\documentclass[
    reprint,
    nofootinbib,
    amsmath,
    amssymb,
    aps,
    prl
]{revtex4-2}

\usepackage{physics}
\usepackage{dsfont}
\usepackage{bm}
\usepackage[scaled=0.92]{helvet}
\usepackage{graphicx}
\usepackage{xcolor}
\usepackage{hyperref}

\providecommand{\argmin}{\operatornamewithlimits{argmin}}

\newcommand{\OH}{\mathbf{O}_H}
\newcommand{\OS}{\mathbf{O}_S}

\makeatletter
\@ifundefined{h}
  {\newcommand{\h}{\mathbf{H}}}
  {\renewcommand{\h}{\mathbf{H}}}
\makeatother

\renewcommand{\eqref}[1]{Eq.~(\ref{#1})}
\renewcommand{\[}{\begin{equation}}
\renewcommand{\]}{\end{equation}}

\newcommand{\covarsp}[3]{(#1,#2)_{\mathrm{covar}(#3)}}
\newcommand{\normsp}[2]{\ensuremath{\left\lVert#1\right\rVert_{#2}}}

\newcommand{\genH}{\bm{\mathcal{L}}}

\begin{document}

\title{State-adapted generalized mean-field projections for Pauli propagation of Heisenberg dynamics}

\author{Federico Tom{\'a}s B. P{\'e}rez}
\email{ftbperez.quantum@gmail.com}
\affiliation{Institut für Theoretische Physik und IQST, Albert-Einstein-Allee 11, Universität Ulm, D-89081 Ulm, Germany}
\affiliation{IFLP - CONICET, Departamento de Física, Facultad de Ciencias Exactas, Universidad Nacional de La Plata, C.C. 67, 1900 La Plata, Argentina}

\author{Gabriela W{\'o}jtowicz}
\affiliation{Institut für Theoretische Physik, Albert-Einstein-Allee 11, Universität Ulm, D-89081 Ulm, Germany}

\author{Martin B. Plenio}
\affiliation{Institut für Theoretische Physik, Albert-Einstein-Allee 11, Universität Ulm, D-89081 Ulm, Germany}
\affiliation{Center for Integrated Quantum Science and Technology (IQST), 89081 Ulm, Germany}

\date{September 11, 2026}

\begin{abstract}
The Heisenberg picture can make quantum many-body simulation efficient when evolved observables admit compact operator representation within low-dimensional structures.
Conventional Pauli-string propagation and truncation techniques exploit this structure, but in coherent Hamiltonian dynamics their error control is often heuristic and their stability can be poor.
We introduce a state-adapted Krylov framework based on geometric generalized mean-field projections of many-body observables onto low-body operator subspaces.
The resulting dynamics approximate expectation values, do not extend spatial support beyond that prescribed by Lieb--Robinson bounds, and compress high-body correlations onto their state-relevant low-body representatives rather than simply discarding them.
We derive necessary operator-entanglement obstructions to low-body representation and conditional sufficient bounds on representation and dynamical errors involving nonstabilizerness and controlled high-body tails.
Numerical benchmarks show improved, stable finite-\(m\) hierarchies and simulations on large three-dimensional lattices, establishing a scalable state-adapted alternative to conventional Heisenberg-picture weight-truncation of Pauli strings.
\end{abstract}

\maketitle

\paragraph{Introduction}

Many-body quantum dynamics produce non-equilibrium phenomena increasingly accessible to programmable quantum processors and analogue simulators, while classical simulation remains indispensable for benchmarking and comparison with experiments~\cite{Georgescu_2014,King_2025,Vetter_2026}.
The central difficulty is the exponential Hilbert-space dimension, together with the interaction-driven spreading of correlations and increasingly high-body operator components.
Special structures evade this growth, including product-state-preserving one-body dynamics, Gaussian quadratic algebras, and Clifford circuits, which preserve Pauli sparsity despite increasing operator weight~\cite{marcinkiewiez_sur_1939,rajagopal_generalizations_1974,gottesman1998,clifford_pert_theo}.
Generic interactions break these closures, motivating systematically improvable low-complexity representations of the resulting correlations.

When only expectation values of selected local observables are required, it is natural to work in the Heisenberg picture,
\begin{equation}
\dot{\bf O}_H(t)=\genH{\bf O}_H(t),
\qquad
\langle{\bf O}(t)\rangle=\Tr[\sigma_0{\bf O}_H(t)],
\label{eq:heisenberg_eom}
\end{equation}
where \(\genH\) is the Heisenberg generator, \(\sigma_0\) is the initial state, and \({\bf O}_H(t_0)={\bf O}_S\) is the observable in the Schr\"odinger picture.
This replaces full-state 
simulation by the propagation of observables of interest, although their representation can itself become non-sparse.

For local generators, Lieb--Robinson (LR) bounds constrain the spatial growth of initially local observables~\cite{lieb1972,eisert2010}, while operator-space entanglement and magic, a measure of non-stabilizerness, characterize complementary aspects of the complexity generated within the resulting causal region~\cite{Prosen2007,dowling2025_magic}.

Tensor-network (TN) methods compress observables through low-rank tensor factorizations whose bond dimension controls the retained intersite correlations~\cite{bridgeman_hand-waving_2017,orus_tensor_2019}.
For a specified bipartition and norm, singular-value truncation gives an optimal local low-rank approximation, but maintaining accuracy in generic real-time dynamics generally requires increasing bond dimension as operator-space entanglement grows~\cite{hartmann2009,clark_exact_2010,muller-hermes_tensor_2012,eisert2010}.
These methods are particularly effective in one dimension, while higher-dimensional contractions are substantially more expensive~\cite{vidal2003,eisert2010,King_2025}.

Pauli-propagation (PP) methods provide a complementary representation of Heisenberg-picture dynamics.
A Heisenberg-evolved observable is expanded as
\begin{equation}
{\bf O}_H(t)
=
\sum_P c_P(t)P,
\qquad
P\in\{I,X,Y,Z\}^{\otimes N},
\end{equation}
where \(P\) denotes a Pauli string and \(c_P(t)\) its time-dependent coefficient.
The complete \(4^N\)-element Pauli basis represents any \({\bf O}_H(t)\) exactly and admits fermionic and Majorana analogues~\cite{zoee_Fermionic,zoee_majorana}.
The computational cost of Pauli propagation is governed by both the number and complexity of the retained Pauli strings, motivating top-\(K\), Pauli-path, coefficient-, weight-, and more recently state-informed truncation criteria, such as X-truncated sparse Pauli dynamics (xSPD) for computational-basis initial states~\cite{shao2026_topK,Angrisani_2025,Angrisani_2026,GGarcia_Pauli_path_2025,clifford_pert_theo,Loizeau_2025,pauli-qite,Vetter_2026,Li_2026,xSPD2026}.
For Pauli-path approaches in particular, rigorous polynomial-time guarantees are known in relevant noisy-circuit settings, where noise suppresses the contribution of sufficiently long or high-weight paths~\cite{Angrisani_2026,schuster2024}.
For generic coherent Hamiltonian dynamics, however, this suppression mechanism is absent, and such guarantees do not translate directly to a general truncation of the evolving Pauli expansion.
Small-amplitude many-body components can accumulate or subsequently feed back into retained sectors, so practical truncation criteria may provide substantially weaker finite-time error control.

In this Letter, we introduce a general state-adapted alternative to standard weight-based Pauli-string truncation and state-informed variants such as xSPD~\cite{xSPD2026}.
Rather than selecting components according to a prescribed Pauli-string criterion, we project onto the at-most-\(m\)-body subspace using the covariance geometry induced by the initial state~\cite{perez_quantum_2024,Perez_2026,Perez_long_paper_2026}.
The resulting static projection $\pi_m^{\sigma_0}$ preserves the target expectation value with respect to $\sigma_0$.

Generic Hilbert--Schmidt (HS) weight truncation has no analogous expectation-preservation property for a non-maximally-mixed reference state.
Neither projection alone, however, guarantees spectral admissibility or finite-time accuracy of the corresponding restricted dynamics.

We combine this projection with an adaptive Krylov-like Hierarchical-Basis evolution~\cite{saad_krylov,adolfo_del_campo_rev,perez_quantum_2024}.
Its error separates into a representational contribution arising from restriction to fixed body order \(m\) and a dynamical contribution generated by the resulting restricted evolution.
Operator-space entanglement provides a necessary obstruction to low-body HS representation, while sufficient accuracy bounds additionally require control of discarded high-body tails.
For local Hamiltonians, LR locality restricts the available fixed-\(m\) operator space to polynomial growth in time, while the adaptive construction explores only its dynamically relevant portion.

We formulate the method explicitly for spin-\(\frac12\) systems and product reference states, with extensions to arbitrary finite-dimensional systems and correlated input states.
Numerical benchmarks compare the state-adapted hierarchy with standard weight-based Pauli truncation and TN calculations for three-dimensional transverse-field Ising dynamics, including lattices of up to \(N=90\) spins.

\paragraph{Generalized mean-field projections for Heisenberg-picture simulation}

\begin{figure}
    \centering
    \includegraphics[width=1\linewidth]{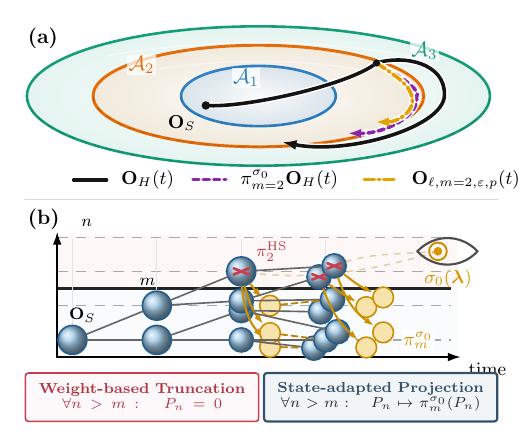}
\caption{
\textbf{State-adapted \(m\)-body Heisenberg dynamics}.
(a) Nested sectors \(\mathcal A_1\subset\mathcal A_2\subset\mathcal A_3 \subset \cdots\) and trajectories for an initially one-body observable \({\bf O}_S\): exact but intractable evolution \({\bf O}_H(t)\), intractable post-facto projection \(\pi_{m=2}^{\sigma_0}{\bf O}_H(t)\), and adaptive restricted flow \({\bf O}_{\ell,m=2,\varepsilon,p}(t)\).
(b) Growth in complexity of the restricted flow in time and how the choice of geometry matters. 
HS weight truncation truncates observable components with weight \(>2\), while \(\pi_m^{\sigma_0}\) compresses their state-relevant lower-body content back into \(\mathcal A_m\).
}
\label{fig:pedagological}
\end{figure}

Standard PP truncations select components through state-independent top-$K$, path, coefficient, or weight criteria
~\cite{shao2026_topK,Angrisani_2025,Angrisani_2026,GGarcia_Pauli_path_2025,clifford_pert_theo,Loizeau_2025,pauli-qite,Vetter_2026,Li_2026} with progress in state-dependent ones such as \cite{xSPD2026}.
We instead project onto nested at-most-\(m\)-body sectors ${\cal A}_0\subset{\cal A}_1\subset{\cal A}_2\subset\cdots$, with \(I\in{\cal A}_0\), using a geometry adapted to the initial state.
Unlike cumulant- or BBGKY-type closures~\cite{Plankensteiner_2022}, the construction is defined geometrically as a metric projection onto the fixed retained operator space \({\cal A}_m\).

We endow the real vector space of Hermitian observables ${\cal A}$ with the (uncentered) covariance bilinear form, with reference and initial state $\sigma_0$ which we restrict to be product states \cite{perez_quantum_2024, Perez_2026, Perez_long_paper_2026}
\begin{equation}
({\bf A},{\bf B})_{\rm covar(\sigma_0)} =
\Tr\!\left[
\sigma_0\frac{\{{\bf A},{\bf B}\}}{2}
\right].
\label{eq:cov}
\end{equation}
For full-rank $\sigma_0$, ~\eqref{eq:cov} is positive definite and defines a scalar product and covariance norm.
For rank-deficient $\sigma_0$, it defines a seminorm on ${\cal A}$ and a scalar product on the quotient space ${\cal A}/{\cal N}_{\sigma_0}$, where ${\cal N}_{\sigma_0}=\{{\bf A}:\|{\bf A}\|_{\rm covar(\sigma_0)}=0\}$ is the covariance-null space.
For a pure reference state $\sigma_0=\ket{\Psi}\!\!\bra{\Psi}$, a Hermitian observable belongs to ${\cal N}_{\sigma_0}$ exactly when ${\bf A}\ket{\Psi}=0$.
Expectation values are recovered as overlaps with the identity, $\langle{\bf O}\rangle_{\sigma_0}=({\bf O},{\bf 1})_{\rm covar(\sigma_0)}$, while $\sigma_0\propto I$ recovers the Hilbert--Schmidt (HS) scalar product.
We use product reference states in the main construction; correlated references and other general properties of the geometry are discussed in the Supplemental Information (SI).

For a product state $\sigma_0=\otimes_{i=1}^N \sigma_i$ and a product observable ${\bf P}=\bigotimes_{i\in X}{\bf p}_i$ with support $X \subset \Lambda$, let $\mu_i=\Tr(\sigma_i{\bf p}_i)$ and ${\bf r}_i={\bf p}_i-\mu_i I_i$ be the centered local observables, so that $\Tr(\sigma_i{\bf r}_i)=0$.
The state-adapted $m$-body projection $\pi_m^{\sigma_0}$ is
\begin{equation}
\pi_m^{\sigma_0}({\bf P}) =
\sum_{\substack{S\subseteq X\\ |S|\leq m}}
\left(
\prod_{i\in X\setminus S}\mu_i
\right)
\bigotimes_{i\in S}{\bf r}_i.
\label{eq:m-body-proj}
\end{equation}
The map extends to arbitrary observables by linearity.
For full-rank $\sigma_0$, ~\eqref{eq:m-body-proj} is the unique observable covariance-orthogonal projection onto ${\cal A}_m$ and solves the following variational problem
\begin{equation}
\pi_m^{\sigma_0}{\bf O}
=
\argmin_{{\bf A}\in{\cal A}_m}
\|{\bf O}-{\bf A}\|_{\rm covar(\sigma_0)}.
\label{eq:fullrank_variational_projection}
\end{equation}
For rank-deficient $\sigma_0$, the variational problem determines only an equivalence class in ${\cal A}_m$, or equivalently a unique element of the quotient space ${\cal A}_m/({\cal A}_m\cap{\cal N}_{\sigma_0})$, while~\eqref{eq:m-body-proj} selects the definite representative in ${\cal A}_m$ used below.
The minimizing representative obeys,
\begin{equation}
\left(
{\bf B},
{\bf O}-\pi_m^{\sigma_0}{\bf O}
\right)_{\rm covar(\sigma_0)} = 0,
\qquad
\forall\,{\bf B}\in{\cal A}_m .
\label{eq:projection-orthogonality}
\end{equation}
Thus, unlike HS weight truncation, which discards an entire Pauli string once its weight exceeds $m$, the state-adapted projection retains all of its centered components within body order up to $m$.
Importantly, the projection preserves the expectation value with respect to $\sigma_0$ and, for product initial states, does not enlarge spatial support.

Hence, the restricted Heisenberg dynamics (covariance dynamics for short) is defined by
\begin{equation}
\dot{\bf O}_m(t) =
\bm{\mathcal L}_m{\bf O}_m(t),
\quad
{\bf O}_m(t_0)=\pi_m^{\sigma_0}{\bf O}_S,
\label{eq:restricted_variational}
\end{equation}
where $\pi_m^{\sigma_0}\bm{\mathcal L}\equiv\bm{\mathcal L}_m$ is the projected generator of the Heisenberg dynamics and ${\bf O}_m(t)$ is the $m$-body flow, which we will later approximate further via a Krylov-like evolution.
For full-rank reference states, ~\eqref{eq:restricted_variational} is equivalently the instantaneous best approximation to the exact Heisenberg tangent generated from ${\bf O}_m(t)$, in the covariance geometry.
For rank-deficient reference states, the variational problem determines the same covariance equivalence class, while \eqref{eq:m-body-proj} fixes the chosen representative propagated by \eqref{eq:restricted_variational}.
The restricted flow is distinct from the post-facto projection $\pi_m^{\sigma_0}{\bf O}_H(t)$, which requires complete knowledge of $\OH(t)$, since the exact trajectory may leave (and return to) ${\cal A}_m$ before being projected, whereas ${\bf O}_m(t)$ remains in ${\cal A}_m$ at all times.

{Rank deficiency raises an alternative, covariance-specific dynamical compatibility question.
When \([\sigma_0,{\bf H}]\neq0\), a covariance-null direction need not remain null under the restricted dynamics.
At first order, retained covariance-equivalent representatives generate the same tangent only if $\pi_m^{\sigma_0}\genH{\bf D}$ remains covariance-null for their difference \({\bf D}\in{\cal A}_m\cap{\cal N}_{\sigma_0}\) i.e. if 
\begin{equation}
    \normsp{\pi_m^{\sigma_0}\genH{\bf D}}{\rm covar(\sigma_0)}=0.
\end{equation}
Failure identifies a potentially broken aliasing channel, but not necessarily a failing trajectory, since the offending direction need not be dynamically populated.
This first-order criterion therefore diagnoses representative compatibility in the rank-deficient covariance quotient rather than spectral admissibility of truncated Heisenberg dynamics in general, which can fail also for state-independent HS truncations.
The stronger condition that \( \pi_m^{\sigma_0}\genH ({\cal A}_m \cap {\cal N}_{\sigma_0})\subseteq{\cal N}_{\sigma_0}\) is sufficient for all retained covariance-null directions but is not required for a particular trajectory.
This point and the distinction from generic truncation errors are analyzed further in the SI and Ref.~\cite{Perez_long_paper_2026}.}

We approximate the \(m\)-body flow \({\bf O}_m(t)\) of~\eqref{eq:restricted_variational} with an adaptive at-most $(\ell+1)$-dimensional Krylov-like Hierarchical-Basis (HB) construction~\cite{perez_quantum_2024,saad_krylov,adolfo_del_campo_rev}, with a built-in diagnosis for the triggering of the adaptive Krylov-like charts.
We denote the resulting approximation by \({\bf O}_{\ell,m,\varepsilon,p}(t)\), where \(m\) fixes the retained body order, \(\ell+1\) the dimension of each Krylov-like Hierarchical Basis (HB) chart, \(p\) the number of highest-index HB directions entering the reconstruction diagnostic, and \(\varepsilon\) the threshold that triggers a chart reconstruction.
Within the \(n\)th chart, triggered at \(T_n\),
\begin{equation}
{\bf O}_{\ell,m,\varepsilon,p}(t)=
\sum_{\alpha=0}^{\ell}
\psi_\alpha^{(n)}(t)\,{\bf b}_\alpha^{(n)},
\qquad
{\bf b}_{\alpha+1}^{(n)}=
\bm{\mathcal L}_m{\bf b}_\alpha^{(n)},
\label{eq:hb_ansatz}
\end{equation}
with ${\bf b}_0^{(n)} = {\bf O}_{\ell,m,\varepsilon,p}(T_n)$.
The $n$-chart obey coordinates 
\begin{equation}
{\cal G}^{(n)}\dot{\vec\psi}^{(n)}= {\cal H}^{(n)}
\vec\psi^{(n)} ,
\label{eq:hb_galerkin}
\end{equation}
where the Gram matrix is defined as,
\begin{equation}
{\cal G}_{ij}^{(n)} = \covarsp{{\bf b}_i^{(n)}}{{\bf b}_j^{(n)}}{\sigma_0},
\label{eq:hb_gram}
\end{equation}
and the corresponding reduced generator matrix is
\begin{equation}
{\cal H}_{ij}^{(n)} =
\covarsp{{\bf b}_i^{(n)}}{
\bm{\mathcal L}_m{\bf b}_j^{(n)}
}{\sigma_0}.
\label{eq:hb_generator}
\end{equation}
Each chart provides the local HB approximation over an interval \([T_n,T_{n+1})\).
The next reconstruction time \(T_{n+1}\) is chosen when the tail diagnostic associated with the \(p\) highest-index HB directions reaches the prescribed threshold \(\varepsilon\).
The chart is then rebuilt around the current approximate observable ${\bf b}_0^{(n+1)} = {\bf O}_{\ell,m,\varepsilon,p}(T_{n+1})$.
The explicit reconstruction diagnostic and practical numerical solution of \eqref{eq:hb_galerkin} are given in the SI and discussed further in Ref.~\cite{Perez_long_paper_2026}.

Note that
\begin{equation}
\hspace{-5pt}
\Tr\left[\sigma_0\OH(t)\right] = 
\Tr\left[\sigma_0\pi_m^{\sigma_0}\OH(t)\right]
\approx \Tr\left[\sigma_0{\bf O}_{\ell,m,\varepsilon,p}(t)\right],
\label{eq:expectation_approximation}
\end{equation}
where the approximation arises solely from replacing the post-facto projected trajectory by the restricted adaptive flow.
Moreover,
$|\Tr[\sigma_0\delta{\bf O}]|
\le \|\delta{\bf O}\|_{\rm covar(\sigma_0)}$,
with $\delta{\bf O}=\OH-{\bf O}_{\ell,m,\varepsilon,p}$.

The latter covariance error source separates exactly into representation and leakage contributions,
\begin{equation}
\left\|
\OH-{\bf O}_{\ell,m,\varepsilon,p}
\right\|_{\rm covar(\sigma_0)}^2 = 
\left[\Delta_{\rm rep}^{(m)}\right]^2 + 
\left[\Delta_{\rm leak}^{(\ell,m,\varepsilon,p)}\right]^2,
\label{eq:error_decomposition}
\end{equation}
where the error components read
\begin{align*}
\Delta_{\rm rep}^{(m)}&=
\left\|
(1-\pi_m^{\sigma_0})\OH
\right\|_{\rm covar(\sigma_0)},\\
\Delta_{\rm leak}^{(\ell,m,\varepsilon,p)}
&=
\left\|
\pi_m^{\sigma_0}\OH
-
{\bf O}_{\ell,m,\varepsilon,p}
\right\|_{\rm covar(\sigma_0)}.
\end{align*}
The equality follows because the representation residual
$(1-\pi_m^{\sigma_0})\OH(t)$ is covariance-orthogonal to ${\cal A}_m$, while the leakage term belongs to ${\cal A}_m$.
Operator-space entanglement entropy (OSEE) provides a necessary obstruction to low-body representation of the exact dynamics in HS norm, while sufficient bounds on covariance representation and restricted-flow errors additionally require control of the discarded high-body tail.
The corresponding results, including the role of Pauli-$\ell_1$ nonstabilizerness in the sufficient estimates~\cite{dowling2025_magic}, are given in the SI and Ref.~\cite{Perez_long_paper_2026}.

\paragraph{Cost scaling}

For local Hamiltonians, LR bounds confine the evolution of an initially local observable, up to exponentially small tails, to a region containing ${\cal O}(t^d)$ sites \cite{lieb1972, eisert2010}.
At fixed $m$, the corresponding at-most-$m$-body Pauli space can grow as 
\begin{equation}
{\rm Cost}(t)= {\cal O}\!\left(t^{dm}\right).
\label{eq:complexity}
\end{equation}
Conditionally on the adaptive reconstruction density remaining bounded, summing this retained-space count over the charts gives the conditional cumulative retained-space estimate ${\rm Cum. \,Cost}(t)= {\cal O}\!\left(t^{dm+1}\right).$
The assumptions behind~\eqref{eq:complexity} are stated explicitly in the SI and the estimate should not be interpreted as a wall-time bound.
Note that HS-weight truncation employs the same conservative retained-space counting. 
The difference between covariance and traditional HS-weight truncation lies in the state-adapted projection, through which low-body information is preserved~\cite{perez_quantum_2024,Perez_long_paper_2026}.
The projection itself is fixed by \(m\) and the reference state, while the operator directions dynamically populated within the retained sector depend
on the Hamiltonian generator.

\paragraph{Results}
\label{sec:results}
For the TFIM,
\begin{equation}
    \h_{\mathrm{TFIM}}
    =
    J\sum_{\langle ij\rangle} Z_iZ_j
    +
    h\sum_i X_i ,
\label{eq:tfim_hamiltonian}
\end{equation}
where \(\langle ij\rangle\) denotes nearest-neighbor pairs on the rectangular lattice, we first consider a \(d=3\), \(3\times3\times2\), \(N=18\) lattice with \(J=1\), \(h/J=0.75\), and the fully \(z\)-polarized reference state
\begin{equation}
    \sigma_0
    =
    \ket{\uparrow}\bra{\uparrow}^{\otimes N}.
\label{eq:tfim_initial_state}
\end{equation}
We propagate the corner observable \(\OS=Z_0\) over \(tJ\in[0,1.215]\), comparable to Ref.~\cite{Angrisani_2026}.
The pure reference is rank deficient and therefore probes the regime in which covariance-null directions may become dynamically activated, without such activation by itself implying failure of a particular trajectory.

Exact dynamics are obtained by full-Hilbert-space Schr\"odinger-picture propagation with matrix-free propagation.
We compare covariance and HS adaptive restricted evolutions at \(m=1,2,3\) with an independent TN hierarchy obtained from Heisenberg-picture two-site TDVP--MPO evolution implemented in \textsc{TeNPy}~\cite{tenpy}.
For PP we fix \(\ell=5\), \(p=2\), and \(\varepsilon=10^{-10}\), such that at fixed \(m\) the two hierarchies differ only through the geometry and associated projection.
The TN calculations use \(\chi=8,16,32,64,128\) and one Krylov-enrichment direction, \(d_{\rm Krylov}=1\).
Implementation, numerical tolerances, and hardware details are reported in the SI.

\begin{figure}[t]
    \centering
    \includegraphics[width=1\linewidth]{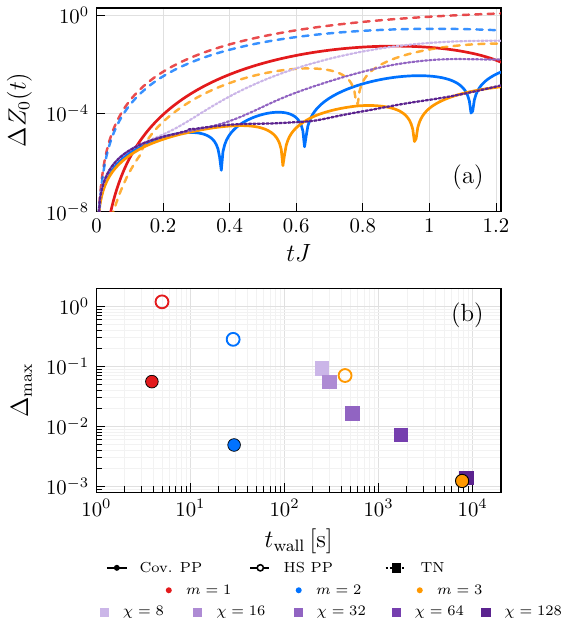}
    \caption{
    \textbf{Accuracy and computational cost for the \(3\times3\times2\) TFIM benchmark.}
    (a) Instantaneous absolute error
    \(\Delta Z_0(t)=
    |\langle Z_0(t)\rangle_{\rm approx}
    -
    \langle Z_0(t)\rangle_{\rm exact}|\)
    for covariance and HS adaptive PP at \(m=1,2,3\), together with representative TN results at \(\chi=8,32,128\).
    (b) Maximum instantaneous error
    \(\Delta_{\max}
    =
    \max_{tJ\in[0,1.215]}\Delta Z_0(t)\)
    versus measured end-to-end wall time, including the full TN hierarchy \(\chi=8,16,32,64,128\).
    The \(\chi=16,64\) trajectories are omitted from panel (a) for clarity.
    }
    \label{fig:covariance-advantage-single}
\end{figure}

Figure~\ref{fig:covariance-advantage-single}(a) isolates the effect of the projection geometry at fixed \(m\).
Both PP hierarchies evolve within the same retained space \({\cal A}_m\), but components generated outside it are treated differently: \(\pi_m^{\sigma_0}\) compresses their state-relevant low-body content according to~\eqref{eq:m-body-proj}, whereas HS weight truncation discards components above body order \(m\).
Despite isolated early-time crossings, covariance PP yields a substantially smaller error envelope and \(\Delta_{\max}\) at every displayed \(m\).
Specifically, \(\Delta_{\max}\) decreases from approximately \(5\times10^{-2}\) at \(m=1\) to \(5\times10^{-3}\) at \(m=2\) and the \(10^{-3}\) scale at \(m=3\), while HS remains respectively at approximately \(10^{0}\), \(10^{-1}\), and \(10^{-2}\).
The independent TN hierarchy likewise approaches the exact trajectory as \(\chi\) is increased {and no expectation-range violation is observed for the displayed HS trajectories.}

Figure~\ref{fig:covariance-advantage-single}(b) shows the corresponding accuracy--cost tradeoff.
At \(m=1,2\), covariance and HS PP have comparable wall times, while covariance reduces \(\Delta_{\max}\) by roughly one to two orders of magnitude.
At \(m=3\), covariance PP becomes substantially denser and more expensive, but also gives the smallest maximum error among the displayed approximations.
The TN hierarchy reaches progressively lower errors with increasing \(\chi\), providing an independent accuracy--cost sequence based on a different representation and propagation scheme.
Since wall time is implementation and hardware dependent, the results should be interpreted as an end-to-end benchmark of the present implementations rather than as an asymptotic complexity statement.

To characterize the PP representation independently of wall time, we use the instantaneous number \({\rm NNZ}(T_n)\) of nonzero Pauli coefficients in the last operator of each HB chart, with coefficients below \(\epsilon_{\rm nnz}=10^{-12}\) ignored.
This quantity measures the sparse operator content actually generated by the adaptive trajectory, rather than the formal dimension of \({\cal A}_m\), and is therefore a representation-level sparsity diagnostic rather than an operation count.
The corresponding observable trajectories, precise NNZ definition, and comparison between \(\Delta_{\max}\) and realized PP representation size are given in the SI.
A complementary analysis of a different family of initial states and a different Hamiltonian is presented in Ref.~\cite{Perez_long_paper_2026}, which also contains an in-depth discussion of dynamical incompatibility.

To probe the large-system regime, we consider the same dynamics on a \(6\times5\times3\), \(N=90\) lattice, retaining \(J=1\), \(h/J=0.75\), the initial state~\eqref{eq:tfim_initial_state}, and the same PP parameters.
Exact propagation is no longer available, so agreement between successive \(m\) provides an internal convergence diagnostic, complemented by the independent TN hierarchy in bond-dimension $\chi$.

\begin{figure}[t]
\centering
\includegraphics[width=1\linewidth]{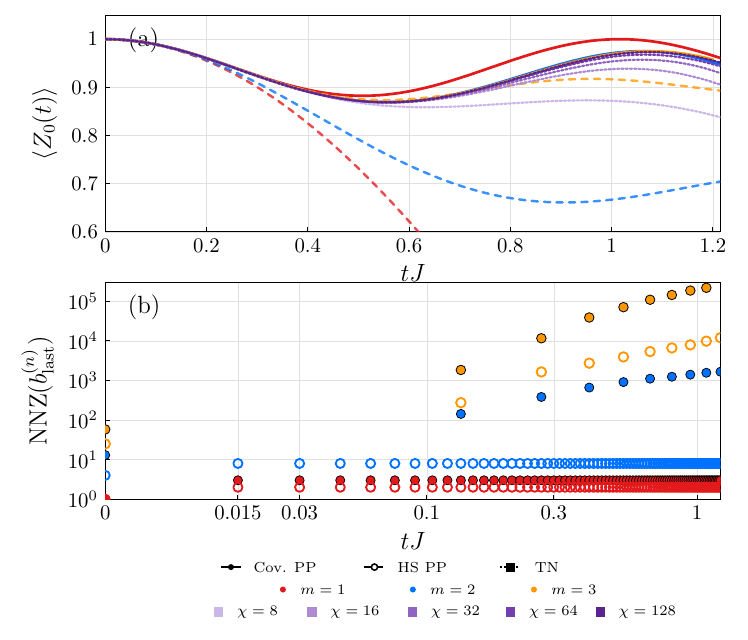}
\caption{
\textbf{Three-dimensional \(N=90\) TFIM benchmark.}
(a) Corner polarization \(\langle Z_0(t)\rangle\) for covariance PP (solid), HS PP (dashed), and the available TN hierarchy (dotted).
(b) Instantaneous sparse content \(\operatorname{nnz}({\bf b}_{\ell}^{(n)})\) at reconstruction times \(T_n\), plotted in a shifted x axis, $x \mapsto x+\zeta$ with $\zeta = 0.075$.
Filled and open markers denote covariance and HS PP, respectively.
Data use \(\ell=5\), \(p=2\), and \(\varepsilon=10^{-10}\).
}
\label{fig:scaling-nnz-inset}
\end{figure}

Figure~\ref{fig:scaling-nnz-inset}(a) shows a markedly cleaner finite-$m$ hierarchy in covariance geometry than in HS.
The covariance $m=2$ and $m=3$ trajectories remain nearly indistinguishable throughout the displayed interval, whereas successive HS orders remain substantially separated, no expectation-range violation is observed in the displayed trajectories.
Although this agreement is not an error certificate, its consistency with the exact $N=18$ benchmark of Fig.~\ref{fig:covariance-advantage-single} and with the independent TN hierarchy provides a useful internal convergence diagnostic.

Figure~\ref{fig:scaling-nnz-inset}(b) shows the corresponding sparse content.
While $m=1$ remains essentially constant in both geometries, higher-order covariance PP generates denser representatives than HS PP, reflecting the additional low-body components retained by~\eqref{eq:m-body-proj}.
Thus, the improved finite-$m$ agreement is accompanied by a broader use of the available low-body sector rather than by a reduction of its realized sparse size.

We quantify this through
\begin{equation}
\mathrm{NNZ}(T_n;\epsilon_{\rm nnz})
=
\operatorname{nnz}\!\left(
{\bf b}^{(n)}_{\ell}
\right),
\label{eq:hb-instantaneous-nnz}
\end{equation}
and fit its finite-time growth as
\(\mathrm{NNZ}(T_n;\epsilon_{\rm nnz})\propto(T_nJ)^\alpha\).

\begin{table}[t]
\caption{
Finite-time effective exponents \(\alpha\) for the \(d=3\), \(6\times5\times3\) TFIM benchmark.
Quoted uncertainties are standard errors of the 
log--log fits; \(\alpha=0\) denotes constant NNZ over the fitted window.
The instantaneous LR retained-space estimate from~\eqref{eq:complexity} is \(\alpha_{\rm wc}=dm\).
Full fit ranges, confidence intervals, numbers of points, and endpoint-sensitivity tests are given in the SI.
}
\label{tab:scaling-fits-main}
\begin{ruledtabular}
\begin{tabular}{cccc}
\(m\) & HS & cov. & \(\alpha_{\rm wc}=3m\) \\
\colrule
\(1\) & \(0\) & \(0\) & \(3\) \\
\(2\) & \(0\) & \(1.105\pm0.063\) & \(6\) \\
\(3\) & \(1.603\pm0.103\) & \(2.319\pm0.115\) & \(9\) \\
\end{tabular}
\end{ruledtabular}
\end{table}

Table~\ref{tab:scaling-fits-main} summarizes the finite-time NNZ growth: HS PP remains constant at \(m=1,2\), while covariance PP produces denser representatives with larger effective exponents, all below the conservative retained-space estimate \(\alpha_{\rm wc}=3m\).
These exponents characterize the sampled time window; fit sensitivity and the relation between sparsity and wall time are discussed in the SI.
Together with the exact \(N=18\) benchmark, the \(N=90\) results support improved finite-\(m\) agreement relative to HS weight truncation, at the cost of denser low-body representations.
This agreement does not certify accuracy or spectral admissibility: either geometry can violate the target spectral bounds, while rank-deficient covariance geometry additionally permits dynamical activation of retained null directions.
The SI gives the corresponding representative-compatibility criterion, with further analysis in Ref.~\cite{Perez_long_paper_2026}.

\paragraph{Conclusions}
In this Letter, we introduced the state-adapted projector \(\pi_m^{\sigma_0}\) as an information-compression principle for Heisenberg-picture simulation.
Unlike weight-based Pauli-string truncation, it retains state-relevant lower-body content and generates a closed hierarchy of restricted dynamics within ${\cal A}_m$.
The resulting low-body description is related in scope to cumulant expansions, BBGKY closures, and Wick-like approximations, but differs in that the static reduction is defined by a state-adapted metric projection rather than by an imposed higher-order closure relation. 
Combined with adaptive HB evolution, the construction has polynomial LR retained-space growth at fixed $m$; under bounded reconstruction density this also yields a polynomial cumulative retained-space estimate.
In the three-dimensional TFIM benchmarks, this state adaptation yields substantially smaller finite-$m$ errors than HS truncation in the exact $N=18$ problem and a markedly cleaner body-order hierarchy at $N=90$.

The framework can be incorporated into existing Pauli-propagation methods by replacing the operator-space geometry and projection rule.
Natural extensions include noisy dynamics, where dissipation may suppress discarded high-body sectors, and fermionic or Majorana formulations based on the corresponding operator algebras.
Determining when noise yields stronger convergence guarantees, and whether state-adapted compression gives comparable advantages for fermionic and Majorana propagation, are promising directions for future work.
Bosonic extensions will be considered separately.

\paragraph{Acknowledgements}
The authors would like to thank Thibaut Lacroix and Piotr Czarnik for interesting discussions.
F.T.B.P. would like to thank Mauricio Matera, Tomás Crosta and Marco Cerezo for interesting discussions and comments on this line of research. 
F.T.B.P acknowledges support from CONICET of Argentina, work supported by CONICET PIP Grant No. 11220200101877CO.
G.~W. acknowledges support from the Alexander von Humboldt Foundation under the Humboldt Research Fellowship. 
G.~W. acknowledges the Financial Support Programmes for Early Career Researchers, Graduate and Professional Training Center, Ulm University for 2025 and 2026/2027. This work was supported by EU-project C-QuENS (Grant No. 101135359) and EU-Project SPINUS (Grant No. 101135699).
The authors acknowledge support by the state of Baden-Württemberg through bwHPC and the German Research Foundation (DFG) through grant no INST 40/575-1 FUGG (JUSTUS 2 cluster).

The authors acknowledge the use of OpenAI ChatGPT, including GPT-5.6 Sol, as an assistive tool during manuscript preparation.
It was used for critical and adversarial examination of selected mathematical arguments, language revision and clarification of scientific text, consistency checks, and
assistance with debugging selected implementation code. 
The scientific ideas, conceptual framework, and principal analytical developments presented here originated with the authors. 
In particular, the state-adapted construction builds on the authors' earlier work on restricted maximum-entropy descriptions, developed prior to the availability of ChatGPT, while the restricted Heisenberg-picture approach underlying the present work was conceived independently of generative-AI assistance.
All AI-assisted suggestions, analytical arguments, numerical procedures, and resulting scientific claims were reviewed and independently verified by the authors, who retain full responsibility for the contents of the manuscript.

\paragraph{Data availability.}
The numerical data supporting the findings of this work, including the data underlying the figures and associated simulation outputs and run metadata, are publicly available in \cite{Perez_PRL_data_2026}.
The custom simulation software used to generate these data is not publicly available at present, as it remains under active development and forms part of ongoing work on the implementation of the method.
Further information regarding the numerical implementation is available from the authors upon reasonable request.

\section{Supplemental Information}

\paragraph{Static properties of the $m$-body projection.}

We first clarify the covariance geometry underlying the state-adapted projection.
Unless stated otherwise, we will restrict our discussion to product-state initial and reference states $\sigma$. 
On the real vector space ${\cal A}$ of Hermitian observables, define the state-dependent covariance bilinear form
\begin{equation}
({\bf A},{\bf B})_{\rm covar(\sigma)}
=
\Tr\!\left[
\sigma\frac{\{{\bf A},{\bf B}\}}{2}
\right]
=
\Re\Tr(\sigma{\bf A}{\bf B}).
\label{eq:covar_sp}
\end{equation}
The associated covariance seminorm is
\begin{equation}
\|{\bf A}\|_{\rm covar(\sigma)}
=
\sqrt{({\bf A},{\bf A})_{\rm covar(\sigma)}}.
\end{equation}
For full-rank $\sigma$,~\eqref{eq:covar_sp} is positive definite and therefore defines a scalar product and norm.
For rank-deficient $\sigma$, nonzero covariance-null observables may exist, and we define
\begin{equation}
{\cal N}_{\sigma} = 
\left\{
{\bf D}\in{\cal A}:
\|{\bf D}\|_{\rm covar(\sigma)}=0
\right\}.
\label{eq:covar_null_SI}
\end{equation}
The covariance form becomes positive definite on the quotient \({\cal A}/{\cal N}_{\sigma}\), whose elements are equivalence classes, i.e. \({\bf A}\simeq_{\mathrm{covar}(\sigma)}{\bf B}\) whenever \({\bf A}-{\bf B}\in{\cal N}_{\sigma}\).
For a pure reference state $\sigma=\ket{\Psi}\!\!\bra{\Psi}$, a Hermitian observable is covariance-null exactly when ${\bf D}\ket{\Psi}=0$.
Expectation values remain overlaps with the identity,
\begin{equation}
\langle{\bf O}\rangle_{\sigma}
=
(I,{\bf O})_{\rm covar(\sigma)}
=
\Tr(\sigma{\bf O}).
\label{eq:expectation_cov_SI}
\end{equation}

Let ${\cal A}_m\subset{\cal A}$ denote the subspace of observables of body order at most $m$.
For full-rank $\sigma$, the covariance projection onto ${\cal A}_m$ is the unique minimizer
\begin{equation}
\pi_m^\sigma{\bf O}
=
\argmin_{{\bf A}\in{\cal A}_m}
\|{\bf O}-{\bf A}\|_{\rm covar(\sigma)}.
\end{equation}
For rank-deficient \(\sigma\), the same variational problem determines an equivalence class \([\pi_m^{\sigma}{\bf O}]\) of minimizing operators in \({\cal A}_m\), or equivalently a unique element of the quotient
\begin{equation}
{\cal A}_m/
\big(
{\cal A}_m\cap{\cal N}_{\sigma}
\big).
\label{eq:retained_quotient_SI}
\end{equation}
Any two minimizing representatives, therefore, differ by an element of ${\cal A}_m\cap{\cal N}_{\sigma}$ and are indistinguishable in the covariance geometry.
The explicit construction below selects one definite representative in ${\cal A}_m$, which is the representative used throughout the projected  and restricted dynamics.

We now specialize to the product reference states used in the main text, $\sigma=\bigotimes_{i\in\Lambda}\sigma_i$.
On each site, the local operator space decomposes algebraically into the identity and the centered subspace,
\begin{equation}
{\cal A}_i
=
{\rm span}\{I_i\}
\oplus
\left\{
{\bf F}_i:
\Tr(\sigma_i{\bf F}_i)=0
\right\}.
\end{equation}
No covariance orthogonality between different centered operators ${\bf F}_i, {\bf F}_j$ on the same support $S$ is required.
For each support $S\subseteq\Lambda$, let ${\cal C}_S$ denote the span of product strings with centered factors on the sites in $S$ and identities elsewhere.
The full observable space, then, decomposes algebraically as
\begin{equation}
{\cal A} =
\bigoplus_{S\subseteq\Lambda}{\cal C}_S,
\qquad
{\cal A}_m
=
\bigoplus_{|S|\leq m}{\cal C}_S.
\label{eq:centered_support_decomposition}
\end{equation}

For product reference states, different centered-support sectors are covariance-orthogonal.
Indeed, if ${\bf A}_S\in{\cal C}_S$ and ${\bf B}_T\in{\cal C}_T$ with $S\neq T$, there exists a site $j\in S\triangle T$ at which one product string contains a centered operator and the other contains the identity.
Factorization of $\sigma_0$ then gives a local factor $\Tr(\sigma_j{\bf F}_j)=0$, and hence
\begin{equation}
({\bf A}_S,{\bf B}_T)_{\rm covar(\sigma_0)}
=
0,
\qquad
S\neq T.
\label{eq:different_support_orthogonality}
\end{equation}
No corresponding orthogonality is assumed within a fixed support sector ${\cal C}_S$, and such same-support orthogonality is not needed for the projection.

Writing an arbitrary observable as
\begin{equation}
{\bf O}
=
\sum_{S\subseteq\Lambda}{\bf O}_S,
\qquad
{\bf O}_S\in{\cal C}_S,
\end{equation}
we define the product-state representative of the $m$-body projection by retaining the centered-support sectors of body order at most $m$,
\begin{equation}
\pi_m^{\sigma_0}{\bf O} = \sum_{|S|\leq m}{\bf O}_S.
\label{eq:centered_sector_projection}
\end{equation}
Its residual is
\begin{equation}
{\bf O}-\pi_m^{\sigma_0}{\bf O}
=
\sum_{|S|>m}{\bf O}_S.
\end{equation}
For any ${\bf B}\in{\cal A}_m$, every retained support $T$ satisfies $|T|\leq m$, whereas every residual support $S$ satisfies $|S|>m$, so necessarily $S\neq T$.
\eqref{eq:different_support_orthogonality} therefore implies
\begin{equation}
\left(
{\bf B},
{\bf O}-\pi_m^{\sigma_0}{\bf O}
\right)_{\rm covar(\sigma_0)}
=
0,
\qquad
\forall\,{\bf B}\in{\cal A}_m.
\label{eq:projection_orthogonality_SI}
\end{equation}
Thus, for full-rank $\sigma_0$, \eqref{eq:centered_sector_projection} is the unique covariance-orthogonal projection onto ${\cal A}_m$.
For rank-deficient $\sigma_0$, it belongs to the minimizing equivalence class selected by the covariance variational problem and fixes the definite representative used in the dynamics.

For a product observable ${\bf P}=\bigotimes_{i\in X}{\bf p}_i$, the same representative has a closed form.
Defining
\begin{equation}
\mu_i
=
\Tr(\sigma_i{\bf p}_i),
\qquad
{\bf r}_i
=
{\bf p}_i-\mu_i I_i,
\label{eq:centered_local_factors_SI}
\end{equation}
one has the exact centered expansion
\begin{equation}
{\bf P}
=
\sum_{S\subseteq X}
\left(
\prod_{i\in X\setminus S}\mu_i
\right)
\bigotimes_{i\in S}{\bf r}_i,
\end{equation}
with identities on $X\setminus S$ implicit.
The $m$-body representative is therefore
\begin{equation}
\pi_m^{\sigma_0}({\bf P})
=
\sum_{\substack{S\subseteq X\\ |S|\leq m}}
\left(
\prod_{i\in X\setminus S}\mu_i
\right)
\bigotimes_{i\in S}{\bf r}_i.
\label{eq:product_projection_SI}
\end{equation}
~\eqref{eq:product_projection_SI} is the explicit prescription employed in the simulations and remains well defined when $\sigma_0$ is rank deficient, fixing a particular gauge and, furthermore, a particular representative in the equivalence class $[\pi_m^{\sigma_0}({\bf P})] \in {\cal A}_m/({\cal A}_m \cap {\cal N}_{\sigma_0})$.

Since $I\in{\cal A}_m$, ~\eqref{eq:projection_orthogonality_SI} implies exact preservation of expectation values,
\begin{equation}
\Tr\!\left[
\sigma_0\pi_m^{\sigma_0}{\bf O}
\right]
=
\Tr(\sigma_0{\bf O}),
\qquad
m\geq0.
\label{eq:expectation_preservation_SI}
\end{equation}
This property holds for the definite representative in ~\eqref{eq:product_projection_SI} and, more generally, for every representative in the same covariance-equivalence class.

For Hermitian observables, the covariance seminorm is bounded by standard state-independent norms,
\begin{equation}
\|{\bf A}\|_{\rm covar(\sigma)}
\leq
\|{\bf A}\|_{\rm op}
\leq
\|{\bf A}\|_{\rm Fro}.
\label{eq:covariance_norm_bounds_SI}
\end{equation}
The covariance projection, therefore, selects low-body information according to the reference state rather than according solely to a state-independent operator norm.

As a simple example, let ${\bf Q}=q_1q_2q_3$ and $a_i=\langle q_i\rangle_{\sigma_0}$.
Its $m=2$ representative is
\begin{equation}
\begin{aligned}
\pi_2^{\sigma_0}({\bf Q})
&=
a_3 q_1q_2
+
a_2 q_1q_3
+
a_1 q_2q_3
\\
&\quad
-
a_2a_3 q_1
-
a_1a_3 q_2
-
a_1a_2 q_3
+
a_1a_2a_3 I.
\end{aligned}
\label{eq:pi2-threebody}
\end{equation}
The lower-body inclusion terms are precisely those generated by the centered expansion and ensure preservation of the expectation value.

For a spin-$1/2$ system of $L$ sites, the dimension of the at-most-$m$-body operator space is
\begin{equation}
\dim({\cal A}_m)
=
\sum_{k=0}^{m}
\binom{L}{k}3^k.
\label{eq:Am_dimension_SI}
\end{equation}
This count depends only on the number of sites and retained body order, while the Hamiltonian, lattice geometry and the initial state determine which of these directions are dynamically generated.

For correlated reference states, the covariance bilinear form, null space, quotient construction, and variational characterization remain unchanged.
However, the factorized centered-support orthogonality used in ~\eqref{eq:different_support_orthogonality}--\eqref{eq:product_projection_SI} is specific to product reference states and does not extend in this form for a generic correlated $\sigma_0$.
When the relevant correlators of $\sigma_0$ can be contracted efficiently, the covariance Gram elements and projected variational problem can still be evaluated without representing the full density operator, but the resulting projection need not preserve the same support-local structure nor maintain the same level of complexity.

\paragraph{Restricted equation of motion}

The expression in~\eqref{eq:restricted_variational} can be understood as the local-in-time closure of a Nakajima-Zwanzig-like master equation for the retained $m$-body sector of the Heisenberg dynamics. 
The purpose of this master equation is to describe how the retained $m$-body operator components evolve when the dynamics generates correlations outside the retained sector and how those may refactor into the description at a later time. 
Its structure is analogous to a Nakajima--Zwanzig equation, but the relevant split is not between a system and a bath but rather between the retained $m$-body operator components and discarded correlations of body order $n\geq m+1$, which otherwise would play the role of an environment.

The exact equation, then, contains both an instantaneous restricted evolution and a memory contribution. 
The latter describes the back-action of discarded high-body components, $n$-body  correlations --- with $n \ge m+1$ --- that are negligible, or compressed away, at a given time $t$, that can evolve in the discarded sector and later feed back into the retained $m$-body dynamics at a later time $t' > t$.
Thus, the restricted Heisenberg equation in~\eqref{eq:restricted_variational} is obtained by neglecting this memory kernel and evolving only with the restricted generator.

Instead of the usual decomposition of the Hamiltonian in terms of the retained system and the bath, we start from considering the projectors $P=\pi_m^{\sigma_0}$ and $Q=1-P$. 
Projecting the exact Heisenberg equation $\dot{\bf O}_H(t) = \bm{\mathcal L} \OH(t)$ gives formally
\begin{align}
    P\dot{{\bf O}}_H (t)&=
    P \bm{\mathcal L} P\,\OH(t)+
    P \bm{\mathcal L} e^{tQ \bm{\mathcal L}} Q \OH(0) \nonumber\\
    &+ 
    \int_0^t ds\,
        P \bm{\mathcal L} e^{(t-s)Q \bm{\mathcal L}}Q \bm{\mathcal L} P\,\OH(s).
\end{align}
For initially retained observables, $Q \OH(0)=0$. 
Dropping the remaining memory kernel yields
\[
    \dot {\bf O}_{m}(t)
    =
    \bm{\mathcal{L}}_m\,{\bf O}_{m}(t), \qquad \bm{\mathcal{L}}_m({\bf A}) = \pi_m \bm{\mathcal{L}}\pi_m{\bf A},
\]
where, in the unitary case, $\genH$ becomes \eqref{eq:restricted_variational} with $\bm{\mathcal{L}}_m({\bf A}) = i \pi_m^{\sigma_0} [{\bf H}, \pi_m^{\sigma_0} {\bf A}]$.
This is therefore a self-consistent evolution inside the retained $m$-body operator subspace, rather than a post-facto truncation of the exact observable $P \OH(t)$.

A natural possible extension would be to combine the retained--discarded correlation split used here with the usual system--bath formulation of open quantum dynamics. 
This setting involves two distinct reductions, the projection $\pi_m^{\sigma_0}$ onto low-body operator components and a system projector ${\cal R}$ selecting the relevant open-system degrees of freedom. 
The interplay between these projections can generate a richer memory structure, with feedback both from discarded high-body correlations and from eliminated bath degrees of freedom. 
In particular, retained $m$-body system--bath correlations may be dynamically weak over long time intervals and re-enter the effective description only perturbatively or at later times. 
A careful derivation of such mixed correlation--bath closures is beyond the scope of this work.

As a final note, we compare the variational principles of the post-facto and restricted observables.
For full-rank $\sigma_0$, the post-facto projection satisfies
\begin{equation}
\label{eq:postfacto_variational}
\pi_m^{\sigma_0}{\bf O}_H(t)
= \argmin_{{\bf A}\in{\cal A}_m}
\left\|
{\bf A}-{\bf O}_H(t)
\right\|_{\mathrm{covar}(\sigma_0)}.
\end{equation}

The fully restricted dynamics satisfies, instead,
\begin{equation}
\dot{\bf O}_{m}(t) = \argmin_{{\bf X}\in{\cal A}_m}
\left\|
{\bf X} - \bm{\mathcal L}{\bf O}_{m}(t)
\right\|_{\mathrm{covar}(\sigma_0)}.
\end{equation}

Equivalently, the post-facto projection is $\pi_m \OH(t) = \pi_m e^{t \genH} \OS$ while the restricted Heisenberg solution is obtained from a true generator ${\genH}_m$ as ${\bf O}_{m}(t) = e^{{\genH_m } t}\OS$.

\paragraph{Relation to hierarchy closures and state-informed truncations.}

Low-correlation-order reductions also arise in generalized cumulant expansions, BBGKY-type closures, and restricted-state-space methods, but through different approximation principles.
Generalized cumulant methods close a hierarchy of expectation-value equations by neglecting cumulants above a prescribed order and expressing the resulting higher moments in terms of retained lower-order moments ~\cite{Plankensteiner_2022}.
Related BBGKY constructions truncate an exact hierarchy of reduced correlation functions or density operators through an additional closure assumption.
State-informed sparse-Pauli methods such as xSPD instead modify the keep/discard criterion according to the initial state ~\cite{xSPD2026}.

The present construction differs in that the static reduction is defined independently of the dynamics as a metric projection of the observable onto ${\cal A}_m$.
No factorization of higher-order correlations or additional Pauli-string selection rule is required to define this map.
Consequently, variational optimality, covariance orthogonality of the residual, preservation of the reference-state expectation value, and nonexpansion of spatial support are properties of the projection itself.
Only subsequently is the exact retained--discarded evolution replaced by a closed dynamics in ${\cal A}_m$, which constitutes a separate dynamical approximation.

\paragraph{Correlated reference states}

The covariance geometry extends directly to correlated reference states, although the explicit site-factorized form of~\eqref{eq:m-body-proj} is then generally lost.
For any reference state \(\sigma_0\),
\begin{equation}
({\bf A},{\bf B})_{\mathrm{covar}(\sigma_0)}
=
\frac{1}{2}
\Tr\!\left[
\sigma_0\{{\bf A},{\bf B}\}
\right], \,\,
\label{eq:app-correlated-covariance}
\end{equation}
and $\|{\bf A}\|_{\mathrm{covar}(\sigma_0)}^2=\Tr(\sigma_0{\bf A}^2)$.
If \({\bf P}_0\) projects onto the support of \(\sigma_0\), then
\begin{equation}
{\cal N}_{\sigma_0}
=
\{
{\bf A}:{\bf A}{\bf P}_0=0
\}.
\label{eq:app-correlated-quotient}
\end{equation}
Thus, the covariance form is an inner product for full-rank \(\sigma_0\) and otherwise becomes one on the corresponding quotient.

For a basis \(\{{\bf Q}_a\}\) of \({\cal A}_m\), the projection is obtained from the variational problem~\eqref{eq:fullrank_variational_projection}.
Defining
\begin{equation}
G_{ab}
=
({\bf Q}_a,{\bf Q}_b)_{\mathrm{covar}(\sigma_0)},
\quad
y_a
=
({\bf Q}_a,{\bf O})_{\mathrm{covar}(\sigma_0)},
\end{equation}
a convenient representative is
\begin{equation}
\pi_m^{\sigma_0}{\bf O}
=
\sum_{a,b}
{\bf Q}_a
(G^+)_{ab}
y_b .
\label{eq:app-correlated-projection}
\end{equation}
For rank-deficient \(\sigma_0\), different solutions can differ only by an element of
\({\cal A}_m\cap{\cal N}_{\sigma_0}\), so the projected equivalence class is unique even when its operator representative is not.
Moreover, since \(I\in{\cal A}_m\),
\begin{equation}
\Tr\!\left[
\sigma_0\pi_m^{\sigma_0}{\bf O}
\right]
=
\Tr(\sigma_0{\bf O}).
\end{equation}
Unlike the product-state construction, however, the required Gram data no longer factorizes site-wise and their evaluation may itself become computationally costly.

\paragraph{Hierarchical Basis representation.}

The restricted Heisenberg equation \eqref{eq:restricted_variational} is solved in a finite adaptive Krylov-like Hierarchical Basis, representation. 
Around an initial time $t_0$, we define the $\ell+1$-dimensional operator subspace
\begin{equation}\label{eq:krylov_basis}
    \mathcal K_\ell(t_0)
    =
    {\rm span}\{ {\bf b}_0,\ldots,{\bf b}_{\ell}\},
\end{equation}
with
\begin{equation}\label{eq:hb_recursion}
    {\bf b}_0 = {\bf O}_{m}(t_0),
    \qquad
    {\bf b}_{\alpha+1}
    =
    \bm{\mathcal L}_m({\bf b}_\alpha).
\end{equation}
Thus, the Hierarchical Basis follows the locally restricted dynamics generated by $\bm{\mathcal L}^{\sigma_0}_m$ rather than spanning the full operator algebra but, due to the nature of the $m$-body projection and its qualities, at the same time, respecting the growing support of the exact dynamics, prescribed by the LR bounds.

Inside this basis the observable is represented as
\begin{equation}
    {\bf O}_{\ell,m}(t)
    =
    \sum_{\alpha=0}^{\ell}
    \psi_\alpha(t)\,{\bf b}_\alpha,
\end{equation}
where the $\psi_\alpha(t)$ are local coordinates associated to the HB chart. 
Substituting this ansatz into the restricted equation and projecting onto the basis gives
\begin{equation}\label{eq:hb_matrix_eom}
    {\cal G}\,\dot{\vec\psi}(t)
    =
    {\cal H}\,\vec\psi(t),
\end{equation}
where
\begin{equation}
    {\cal G}_{\alpha\beta}
    =
    ({\bf b}_\alpha,{\bf b}_\beta)_{\rm covar(\sigma_0)},
    \quad
    {\cal H}_{\alpha\beta}
    =
    ({\bf b}_\alpha,
    \bm{\mathcal L}_m{\bf b}_\beta)_{\rm covar(\sigma_0)} .
\end{equation}
For compatible charts, the minimum-norm coefficient evolution is
\begin{equation}\label{eq:hb_coeff_eom}
\dot{\vec\psi}(t)
=
{\cal G}^{+}{\cal H}\vec\psi(t),
\end{equation}
where ${\cal G}^{+}$ denotes the Moore--Penrose pseudoinverse and fixes the minimum-norm coefficient evolution when the covariance Gram matrix is singular.
The numerical treatment of potentially singular or ill-conditioned Gram systems used in the simulations is specified in the implementation details below.

For a fixed basis, the coefficient evolution is therefore
\begin{equation}\label{eq:hb_solution}
    \vec\psi(t)
    =
    \exp\!\left[
        (t-t_0){\cal G}^{+}{\cal H}
    \right]
    \vec\psi(t_0).
\end{equation}

The parameter $\ell$ controls the local Krylov depth. 
For short time intervals, the first $\ell$ restricted commutators approximate the local exponential propagation and increasing $\ell$ systematically captures higher orders in the local time expansion. 
For local Hamiltonians, LR bounds imply that the relevant operator support grows only within an effective light cone, so a finite $\ell$ can describe the dynamics over a finite time window. 
At longer times the basis, due to the fact that any finite basis inevitably loses track of the dynamics, is rebuilt adaptively around the current operator, producing a sequence of local charts rather than a single global Krylov basis.

This is the practical content of the Hierarchical Basis construction.
The algorithm explores only the dynamically generated, state-relevant part of the retained $m$-body operator sector, instead of manipulating the full space ${\cal A}_m$ at once.

\paragraph{Adaptive reconstruction and full ansatz.}

The finite Hierarchical Basis is used only for a self-determined finite window of time.
Given a chart initialized at time $T_n$, we  construct the local basis
\begin{equation}
    {\bf b}^{(n)}_0
    =
    {\bf O}_{\ell,m,\varepsilon,p}(T_n),
    \quad
    {\bf b}^{(n)}_{\alpha+1}
    =
    \bm{\mathcal L}_m
    \big({\bf b}^{(n)}_\alpha\big),
\end{equation}
where $\alpha=0,\ldots,\ell$. 
For $t\in[T_n,T_{n+1})$, the approximate observable is represented in this local chart as
\begin{equation}\label{eq:local_chart_ansatz}
    {\bf O}_{\ell,m,\varepsilon,p}(t)
    =
    \sum_{\alpha=0}^{\ell}
    \psi^{(n)}_\alpha(t)\,
    {\bf b}^{(n)}_\alpha .
\end{equation}
The chart coefficients obey the finite-dimensional projected equation
\begin{equation}\label{eq:psiEOM}
    {\cal G}^{(n)}\dot{\vec\psi}^{(n)}(t)
    =
    {\cal H}^{(n)}\vec\psi^{(n)}(t),
\end{equation}
where
\begin{align}
    {\cal G}^{(n)}_{\alpha\beta}
    &=
    \big(
        {\bf b}^{(n)}_\alpha,
        {\bf b}^{(n)}_\beta
    \big)_{\rm covar(\sigma_0)},
    \\
    {\cal H}^{(n)}_{\alpha\beta}
    &=
    \big(
        {\bf b}^{(n)}_\alpha,
        \bm{\mathcal L}_m
        {\bf b}^{(n)}_\beta
    \big)_{\rm covar(\sigma_0)} .
\end{align}
Equivalently, we can write
\[
    \dot{\vec\psi}^{(n)}(t)
    =
    \big({\cal G}^{(n)}\big)^+
    {\cal H}^{(n)}
    \vec\psi^{(n)}(t),
\]
with the same formal Moore--Penrose prescription as in~\eqref{eq:hb_coeff_eom}, which provides a formal resolved-support reference.
The production trajectories instead use the distinct finite-ridge generator $ A_r=(G+rI)^{-1}H$ described below.
For an exactly compatible positive-semidefinite system, $A_r\to G^{+}H$ as $r\to 0$, but they are not identical at finite $r$.

Importantly, if ${\cal H}\ker{\cal G}\neq\{0\}$, two coefficient vectors representing the same covariance-equivalence class can generate inequivalent resolved derivatives.
The resulting raw-coordinate evolution remains defined once a representative and ridge prescription are fixed, but it does not define a representative-independent dynamics on the covariance quotient.
This is the coordinate-space form of the dynamical-compatibility problem discussed below and is distinct from generic spectral-bound violations caused by finite-body Heisenberg truncation.

Finite-precision coupling into near-null left directions of ${\cal G}$ is a different numerical issue, since such components can be amplified by the regularized solve.
The production diagnostics therefore distinguish structural quotient incompatibility from ordinary conditioning of the finite HB Gram system.

The adaptive reconstruction is controlled by the weight carried by the tail of the local chart. 
For a fixed tail size $p$, define
\[
    \vec\psi^{(n,p)}(t)
    :=
    \big(
        0,\ldots,0,
        \psi^{(n)}_{\ell-p+1}(t),
        \ldots,
        \psi^{(n)}_{\ell}(t)
    \big)^T .
\]
The tail diagnostic is the normalized Euclidean tail fraction
\begin{equation}\label{eq:partial-sum}
     {\cal K}^{(n)}_p(t)
    =
    \frac{
        \sum_{\alpha=\ell-p+1}^{\ell}
        |\psi^{(n)}_\alpha(t)|^2
    }{
        \sum_{\alpha=0}^{\ell}
        |\psi^{(n)}_\alpha(t)|^2
    } .
\end{equation}
Note that $0 \leq p \leq \ell+1$ and where the preceding definition is valid provided $\vec{\psi} \neq \vec{0}$.
This diagnostic is deliberately defined in coefficient space rather than through the covariance Gram matrix.
Its role is to detect migration of the reduced coordinate vector toward the boundary of the finite HB chart, not to measure a fraction of the covariance norm of the observable.
In a rank-deficient or poorly conditioned covariance geometry, a nonzero tail direction may lie in or near the null space of \(\mathcal G^{(n)}\), so a Gram-weighted tail can assign it vanishing or anomalously small weight and delay necessary reconstruction.
By contrast, \(K_p^{(n)}(t)\) is bounded between zero and one, invariant under an overall rescaling of \(\vec\psi^{(n)}\), and independent of the rank and conditioning of \(\mathcal G^{(n)}\), being therefore much more well suited to ill-defined and poorly-conditioned linear algebra problems. 
It remains a chart-dependent diagnostic under independent rescalings or changes of the HB vectors, but the recursive construction above fixes the coordinate convention used throughout the simulations

The next reconstruction time is chosen as
\begin{equation}\label{eq:adaptive_time}
    T_{n+1}
    =
    \inf
    \left\{
        t>T_n:
        {\cal K}^{(n)}_p(t)\geq\varepsilon
    \right\}.
\end{equation}
At $T_{n+1}$, the evolved operator initializes a new chart,
\[
    {\bf b}^{(n+1)}_0
    =
    {\bf O}_{\ell,m,\varepsilon,p}(T_{n+1}),
\]
and the procedure is repeated. 
The tolerance $\varepsilon$ is therefore an adaptivity parameter controlling when the local basis is rebuilt and, importantly, it should not be interpreted as a certifiable a priori error bound.

The complete ansatz is the piecewise-defined operator
\begin{equation}\label{eq:oleps_def}
    {\bf O}_{\ell,m,\varepsilon,p}(t)
    =
    \sum_{\alpha=0}^{\ell}
    \psi^{(n)}_\alpha(t)\,
    {\bf b}^{(n)}_\alpha,
    \qquad
    t\in[T_n,T_{n+1}) .
\end{equation}
It is controlled by four parameters: the body order $m$, which fixes the retained correlation sector, the Hierarchical Bases depth $\ell$, which fixes the local expansion capacity of each chart, the tolerance $\varepsilon$ and $p$, which both fix the reconstruction criterion.
The semigroup property of the Heisenberg evolution allows the local propagations to be concatenated, while LR locality ensures that each chart explores only the dynamically generated operator directions inside the relevant light cone.

\paragraph{{Spectral admissibility and Hilbert--Schmidt norm control.}}

It is useful to distinguish norm preservation of the operator from preservation of the spectral properties of the observable.

For an initial Pauli observable ${\bf O}_S \in \{I,X,Y,Z\}$, exact unitary Heisenberg evolution gives
\begin{equation}
{\bf O}_H(t) = U^\dagger(t){\bf O}_S U(t),
\quad {\bf O}_H(t)^2=I .
\label{eq:si-exact-pauli-involution}
\end{equation}
Consequently,
\begin{equation}
\normsp{{\bf O}_H(t)}{\rm op}=1,
\quad
\left| \Tr[\rho\,{\bf O}_H(t)]\right| \leq1
\label{eq:si-exact-spectral-bound}
\end{equation}
for every physical state $\rho$.
A truncated Heisenberg propagation is generally not a unitary conjugation of the initial observable and need not preserve \eqref{eq:si-exact-pauli-involution}, as reported in \cite{xSPD2026, Li_2026} and \cite{Angrisani_2026} with provable guarantees for some special circuit propagation found in \cite{Angrisani_2025}.

This issue should be distinguished from ordinary loss of HS norm due to truncation. 
Using the normalized HS scalar product 
\begin{equation}
({\bf A},{\bf B})_{\rm HS} = 2^{-N}\Tr({\bf A}{\bf B}),
\label{eq:si-normalized-hs-product}
\end{equation}
an observable
\begin{equation}
{\bf O} = \sum_P c_P P
\end{equation}
satisfies
\begin{equation}
\normsp{{\bf O}}{\rm HS}^2 = \sum_P |c_P|^2 .
\label{eq:si-hs-pauli-two-norm}
\end{equation}
A discrete coefficient-pruning, such as top-$K$ \cite{shao2026_topK}, or post-step weight-projection \cite{Angrisani_2026} operation removes HS-orthogonal Pauli components and, therefore, decreases the HS-norm whenever a nonzero component is discarded.
This behavior should be distinguished from the continuously restricted fixed-$m$ HS flow considered in \eqref{eq:restricted_variational} and \cite{Perez_long_paper_2026}, where HS-norm preservation is guaranteed and explicit at all times. 

The ideal fixed-body-order restricted HS dynamics considered in \eqref{eq:restricted_variational}, under the appropriate HS projector $\pi_m^{I}$, has a different structure. 
Let $P_m^{\rm HS} = \pi_m^{I}$ be the HS-orthogonal projector onto ${\cal A}_m$ and define
\begin{equation}
\genH_m^{\rm HS} = P_m^{\rm HS}\genH P_m^{\rm HS}.
\label{eq:si-hs-restricted-generator}
\end{equation}
For Hamiltonian Heisenberg evolution, $\genH$ is skew-adjoint in the HS product, while $P_m^{\rm HS}$ is self-adjoint.
Hence
\begin{equation}
\left( \genH_m^{\rm HS}
\right)^\dagger = -\genH_m^{\rm HS},
\label{eq:si-hs-projected-skew}
\end{equation}
and the ideal restricted flow $\dot{\bf O}_m=\genH_m^{\rm HS}{\bf O}_m$
obeys
\begin{equation}
\frac{d}{dt}
\normsp{{\bf O}_m(t)}{\rm HS}^2 = 0 .
\label{eq:si-hs-norm-conservation}
\end{equation}
Thus, loss of the HS norm is not intrinsic to every form of HS Pauli truncation.
Threshold-based pruning and continuous orthogonal projection onto a fixed Pauli-weight sector should therefore be distinguished.
The finite adaptive-HB implementation may introduce additional numerical deviations through the reduced finite-chart solve and its regularization, as discussed in the implementation details, but these are separate from the structural low-body approximation.

Importantly, even exact conservation of~\eqref{eq:si-hs-norm-conservation} does not imply preservation of the operator norm or spectrum.
For example, 
\begin{equation}
{\bf A} = \frac{Z_1+Z_2}{\sqrt{2}}
\label{eq:si-hs-counterexample}
\end{equation}
has
\begin{equation}
\normsp{{\bf A}}{\rm HS}=1, \quad \normsp{{\bf A}}{\rm op}=\sqrt{2},
\end{equation}
but,
\begin{equation}
\bra{00}{{\bf A}}\ket{00} = \sqrt{2}>1.
\end{equation}
Hence, an approximate Pauli observable can exactly preserve the same normalized HS norm as the original Pauli operator, while violating its spectral expectation value-bound and becoming physically inadmissible as a Heisenberg proxy for the exact evolution.

The distinction can also be expressed directly in terms of Pauli coefficients.
For Hermitian ${\bf O}=\sum_Pc_PP$,
\begin{equation}
\normsp{{\bf O}}{\rm HS} = \left(
\sum_P|c_P|^2 \right)^{1/2},
\qquad 
\normsp{{\bf O}}{\rm op}
\leq
\sum_P|c_P|.
\label{eq:si-pauli-l2-l1-op}
\end{equation}
Thus, the coefficient $\ell_1$-norm is not a conserved normalization and need not equal unity.
Rather, it provides a stronger worst-case control on expectation values.
For an approximation error
\begin{equation}
{\bf E} = \widetilde{\bf O}-{\bf O} = \sum_P\delta c_P P,
\end{equation}
one has
\begin{equation}
\left| \Tr(\rho{\bf E}) \right|
\leq
\normsp{{\bf E}}{\rm op} \leq \sum_P|\delta c_P|.
\label{eq:si-l1-expectation-error}
\end{equation}
This difference between coefficient $2$-norm control and the stronger $\ell_1$ information relevant to expectation-value errors has been emphasized in recent analyses of Pauli propagation~\cite{Angrisani_2026}.

Restoring the HS norm after a truncation does not change this conclusion.
For example, rescaling an approximation according to
\begin{equation}
\widetilde{\bf O}
\longrightarrow
\frac{\normsp{{\bf O}_S}{\rm HS}}{
\normsp{\widetilde{\bf O}}{\rm HS}}
\widetilde{\bf O}
\label{eq:si-hs-rescaling}
\end{equation}
can enforce the known HS norm and is used in some Pauli-truncation schemes~\cite{shao2026_topK}, but it does not guarantee $\normsp{\widetilde{\bf O}}{\rm op}\leq1$ or restore the exact spectrum.
We therefore do not regard HS renormalization as a general physicality correction.

A violation of the spectral interval of a bounded target observable is consequently a sufficient witness that a truncated trajectory has failed as an approximation to that observable, but it does not identify the underlying failure mechanism.
Ordinary low-body, state adapted or agnostic, or sparse Pauli truncation can produce such violations without any covariance-null structure.
Rank-deficient covariance dynamics possess the additional possibility that a statically null distinction becomes dynamically covariance resolved.
The projected-activation criterion discussed next diagnoses this particular representative-compatibility problem and should not be interpreted as a general spectral-admissibility criterion.

Conversely, the absence of such a violation for a particular reference  state does not establish spectral admissibility, since the approximate operator may possess eigenvalues outside the target interval that are not resolved by that state.

\paragraph{Dynamical compatibility of the covariance geometry}
\label{sec:dynamical-compatibility}
{The preceding discussion concerns spectral non-preservation that can arise generically under truncated Heisenberg dynamics.
Rank-deficient covariance geometry introduces an additional and distinct structural issue: observables that differ only by a covariance-null direction are identified in the quotient space, although the projected Heisenberg generator need not preserve this static identification.}
This static identification defines representative-independent dynamics only when the unresolved distinction remains unresolved under the relevant projected Heisenberg evolution.
Thus, for an exactly covariance-null observable ${\bf D}\in{\cal N}_{\sigma_0}$, full-space compatibility requires
\begin{equation}
\normsp{ \genH{\bf D} }{ \mathrm{covar}(\sigma_0) } = 0 .
\label{eq:full-space-compat}
\end{equation}
Thus, for unitary dynamics $[\h,\sigma_0]\neq0$ merely permits dynamical activation of covariance-null directions; it does not imply it.

To illustrate this point, consider
\begin{equation}
{\bf A}=X_0X_2,
\qquad
{\bf B}=X_0Z_1X_2,
\qquad
{\bf D}={\bf A}-{\bf B},
\label{eq:compatibility-defect}
\end{equation}
and the homogeneous product reference family
\begin{equation}
\sigma_0(\mu) = \bigotimes_i
\frac{I+\mu Z_i}{2},
\qquad 
0\leq\mu\leq1 .
\label{eq:homogeneous-polarized-reference}
\end{equation}
In the dictionary $\{{\bf A},{\bf B}\}$, the covariance Gram matrix is
\begin{equation}
{\cal G}_{{\bf AB}}^{(\mu)} =
\begin{pmatrix}
1 & \mu \\
\mu & 1
\end{pmatrix}.
\label{eq:compatibility-gram-matrix}
\end{equation}
Introducing the bright and dark combinations \({\bf S}={\bf A}+{\bf B}\) and \({\bf D}={\bf A}-{\bf B}\), their covariance seminorms are
\begin{equation}
\begin{aligned}
\normsp{{\bf S}}{\mathrm{covar}(\sigma_0(\mu))}^2
&= 2(1+\mu),
\\
\normsp{{\bf D}}{\mathrm{covar}(\sigma_0(\mu))}^2
&=
2(1-\mu).
\end{aligned}
\label{eq:compatibility-bright-dark-norms}
\end{equation}
Hence, at the pure-state endpoint $\mu=1$, the Gram matrix has rank one and
\begin{equation}
{\bf A}
\simeq_{\mathrm{covar}(\sigma_0(1))}
{\bf B},
\qquad
{\bf D}
\simeq_{\mathrm{covar}(\sigma_0(1))}
0 .
\label{eq:AB-static-alias}
\end{equation}
The covariance geometry therefore resolves only the bright combination of the corresponding coefficients, while the Moore--Penrose prescription fixes the unresolved dark coordinate to zero.

For the restricted dynamics, let \(P_m(\mu)\equiv\pi_m^{\sigma_0(\mu)}\).
If the covariance-null distinction is itself retained, i.e. ${\bf D}\in{\cal A}_m\cap{\cal N}_{\sigma_0}$, two retained representatives differing by ${\bf D}$ have representative-independent projected derivatives if and only if
\begin{equation}
\normsp{
P_m(\mu)\genH{\bf D} }{
\mathrm{covar}(\sigma_0(\mu))
} = 0.
\label{eq:projected-alias-criterion}
\end{equation}
Full-space compatibility in~\eqref{eq:full-space-compat} implies the absence of covariance-bright projected activation for every \(m\); whenever \({\bf D}\in{\cal A}_m\cap{\cal N}_{\sigma_0}\), this in turn implies representative compatibility of the corresponding retained dynamics. 
The converse need not hold: a covariance-bright full-space image may have no covariance-bright component inside a given retained sector, a question of reachability of the dynamics.

For the TFIM,
\begin{equation}
\h_{\mathrm{TFIM}} = J\sum_i Z_iZ_{i+1} + h\sum_i X_i ,
\label{eq:tfim-compatibility-H}
\end{equation}
the defect in~\eqref{eq:compatibility-defect} satisfies
\begin{equation}
\normsp{{\bf D}}{\mathrm{covar}(\sigma_0(\mu))}=\sqrt{2(1-\mu)}.
\label{eq:defect-norm-mu}
\end{equation}
Since $P_3(\mu)\genH{\bf D}=\genH{\bf D}$, its full Liouvillian image has covariance seminorm
\begin{equation}
\normsp{ \genH{\bf D}}{ \mathrm{covar}(\sigma_0(\mu))} = 2 \sqrt{ h^2 + 4J^2(1-\mu)(1+\mu^2)}.
\label{eq:global-defect-activation}
\end{equation}
Consequently,
\begin{equation}
\normsp{\genH{\bf D}
}{\mathrm{covar}(\sigma_0(1)) } = 2|h| .
\label{eq:global-defect-endpoint}
\end{equation}
Thus, although ${\bf D}$ becomes exactly covariance-null at $\mu=1$, its dynamical image remains covariance-bright, and the corresponding full-space covariance identification is dynamically incompatible.

The extent to which this global incompatibility is visible in a restricted $m$-body sector is determined by the projected Liouvillian images.
For the first three body orders,
\begin{equation}
\begin{aligned}
P_1(\mu)\genH{\bf D} &= 0,\\
P_2(\mu)\genH{\bf D}&= 2J(1-\mu){\bf C},\\ P_3(\mu)\genH{\bf D} &= \genH{\bf D},
\end{aligned}
\label{eq:tfim-projected-images-m123}
\end{equation}
where \({\bf C}=Y_0X_2+X_0Y_2\).
Their covariance seminorms are
\begin{equation}
\begin{aligned}
\normsp{
P_1(\mu)\genH{\bf D}}{
\mathrm{covar}(\sigma_0(\mu))}&= 0,
\\
\normsp{
P_2(\mu)\genH{\bf D}
}{ \mathrm{covar}(\sigma_0(\mu)) }
&= 2|J|(1-\mu)\sqrt{2(1+\mu^2)},
\\
\normsp{ P_3(\mu)\genH{\bf D}}{\mathrm{covar}(\sigma_0(\mu))}
&=2 \sqrt{h^2+4J^2(1-\mu)(1+\mu^2)} .
\end{aligned}
\label{eq:tfim-projected-norms-m123}
\end{equation}
Consequently, at $\mu=1$, \(\normsp{P_1(1)\genH{\bf D}}{\mathrm{covar}(\sigma_0(1))}=0\), \(\normsp{P_2(1)\genH{\bf D}}{\mathrm{covar}(\sigma_0(1))}=0\), and \(\normsp{P_3(1)\genH{\bf D}}{\mathrm{covar}(\sigma_0(1))}=2|h|\).

Since the chosen defect contains a three-body contribution,
\({\bf D}\notin{\cal A}_1\) and \({\bf D}\notin{\cal A}_2\).
The $m=1$ and $m=2$ results therefore do not constitute tests of strict representative compatibility according to~\eqref{eq:projected-alias-criterion}.
Rather, they show that the dynamically generated distinction associated with this globally incompatible null direction is invisible in the $m=1$ and $m=2$ retained sectors at the pure-state endpoint and does not leak into the retained subspaces ${\cal A}_{m=1}$ and ${\cal A}_{m=2}$.

At $m=3$, by contrast, \({\bf D}\in{\cal A}_3\cap{\cal N}_{\sigma_0(1)}\), so the same defect is itself a retained covariance-null direction.
Its nonzero projected activation at $\mu=1$, therefore, violates the representative-compatibility criterion~\eqref{eq:projected-alias-criterion} and identifies a possible retained-null ambiguity.
Nevertheless, this does not by itself predict instability.
Indeed, the $m=3$ covariance TFIM propagation reported in the main text, in \ref{fig:covariance-advantage-single}, remains regular, well-behaved and substantially more accurate than its HS counterpart.
The example therefore illustrates the distinction between global dynamical incompatibility, its visibility within a particular retained sector, and a genuine retained-null representative ambiguity.
Projected activation is a structural diagnostic of the latter: its absence is sufficient at first order for representative-independent and well-defined covariance dynamics along the corresponding retained null direction, whereas nonzero activation does not by itself imply finite-time failure, which additionally requires that direction to be dynamically populated.

\paragraph{Combinatorial Lieb--Robinson retained-space estimate}
\label{app:hb-combinatorial-cost}

In this section, now, we cover the simple counting argument behind the worst-case scaling quoted in \eqref{eq:complexity}.
Let ${\bf P}$ be a product operator with support size $n$.
The projected representative $\pi_m^{\sigma_0}{\bf P}$ keeps all components of body order at most $m$, with weights determined by contractions against the initial product state $\sigma_0$.
If the local contractions with $\sigma_0$ are treated as constant-cost $\mathcal O(1)$ operations, the relevant cost is the number of retained Pauli sub-strings.
For local Hilbert-space dimension $q$, this number is bounded by
\begin{equation}
    N_{\rm proj}(n,m) = 
    \sum_{s=0}^{m}
    \binom{n}{s}
    (q^2-1)^s = {\cal O}(n^m),
\label{eq:supp-proj-cost}
\end{equation}
where  $m,q$ are fixed. 
For spin-$1/2$ systems one has $q=2$, so that $q^2-1=3$ counts the non-identity Pauli labels per site.
Thus, the projection is polynomial in the support size $n$ for fixed $m$, although its prefactor grows exponentially in $m$.
Thus, $N_{\rm proj}(n,m)$ gives the maximum number of $m$-body observable components with shared support in the same $n$ sites.

We now combine this count with Lieb--Robinson locality.
Starting from a local observable $\OS$, the part of the Heisenberg-evolved operator that can influence the dynamics up to time $t$ is contained, up to exponentially small tails, in a causal region $B_{\ell}$ of linear size $B_{\ell} \sim O(v_{\rm LR}t)$.
In spatial dimension $d$, the number of sites in this region obeys
\begin{equation}
    n(t) = {\cal O}\left((v_{\rm LR}t)^d\right) = O(t^d).
\end{equation}
The number of at-most-$m$ Pauli strings available in this region therefore scales as
\begin{equation}
    N_{\leq m}(t) =
    {\cal O}\left(n(t)^m\right) = O(t^{dm}).
\label{eq:single-chart-lr-count}
\end{equation}
This is the worst-case sparse size of any $m$-body projected HB chart at time $t$, within the relevant LR region.

LR locality yields the per-chart retained-space count ${\cal O}(t^{dm})$.
Under the hypothesis that the number of reconstruction times $T_n$ up to time $t$ grows at most linearly with t, the sum over charts is bounded by the time integral of the worst-case chart size. 
Consequently,
\begin{equation}
C_{\rm HB}(t)
\lesssim
\sum_{T_n\leq t} {\cal O}(T_n^{dm})
= {\cal O}\left(\int_0^t d\tau \, \tau^{dm}\right)
= {\cal O}(t^{dm+1}).
\label{eq:hb-cost-lr-estimate}
\end{equation}
This is the combinatorial Lieb--Robinson estimate used in the main text. 
A detailed analysis of the validity of the bounded-reconstruction-frequency hypothesis is important, since a rank-deficient Gram matrix ${\cal G}$ can, in principle, trigger an unbounded number of reconstructions. 
Such an analysis, however, lies beyond the scope of the present work.

The estimate is polynomial in time for fixed $m$ and $\ell$, but its exponent grows linearly with the body order.
It should be understood as a worst-case count of all $m$-body strings allowed by the causal volume, rather than as a prediction that all such strings are dynamically populated, which can be substantially lower due to symmetry-protected sectors, cancellations and integrability.

\paragraph{Representation and leakage errors}

Throughout this section, \(\log\) denotes the natural logarithm.
The covariance error admits the orthogonal decomposition introduced in~\eqref{eq:error_decomposition},
\begin{equation}
\Delta_{\ell,m,\varepsilon,p}^2(t)
=
\left[\Delta_{\rm rep}^{(m)}(t)\right]^2
+
\left[\Delta_{\rm leak}^{(\ell,m,\varepsilon,p)}(t)\right]^2 ,
\label{eq:si-error-decomposition}
\end{equation}
because
\(
(1-\pi_m^{\sigma_0})\OH(t)
\)
is covariance-orthogonal to \({\cal A}_m\), whereas
\(
\pi_m^{\sigma_0}\OH(t)-{\bf O}_{\ell,m,\varepsilon,p}(t)
\in{\cal A}_m
\).
The representation error,
\begin{equation}
\Delta_{\rm rep}^{(m)}(t)
= \normsp{
(1-\pi_m^{\sigma_0})\OH(t)} {\mathrm{covar}(\sigma_0)
},
\label{eq:si-representation-error}
\end{equation}
is therefore a property of the exact observable and the retained \(m\)-body space.
The leakage term contains the additional error generated by replacing the post-facto projected trajectory by the adaptive restricted dynamics.

It is useful to separate this latter contribution once more.
Let
\begin{equation}
{\bf O}_m(t)=
e^{(t-t_0)\pi_m^{\sigma_0}\bm{\mathcal L}\pi_m^{\sigma_0}} \pi_m^{\sigma_0}{\bf O}_S
\label{eq:si-ideal-restricted-flow}
\end{equation}
denote the closed evolution in the retained sector.
Then
\begin{equation}
\Delta_{\rm leak}^{(\ell,m,\varepsilon,p)}
\leq
E_{\rm mem}^{(m)} + E_{\rm HB}^{(\ell,m,\varepsilon,p)},
\label{eq:si-leakage-split}
\end{equation}
where
\begin{align}
E_{\rm mem}^{(m)}(t)
&=\normsp{
\pi_m^{\sigma_0}\OH(t)-{\bf O}_m(t)
}{\mathrm{covar}(\sigma_0)},
\nonumber\\
E_{\rm HB}^{(\ell,m,\varepsilon,p)}(t)
&=
\normsp{{\bf O}_m(t)-{\bf O}_{\ell,m,\varepsilon,p}(t)}{\mathrm{covar}(\sigma_0)}.
\label{eq:si-memory-hb-errors}
\end{align}
Thus, \(E_{\rm mem}^{(m)}\) measures the feedback through the discarded sector inherent to closing the dynamics in \({\cal A}_m\), whereas \(E_{\rm HB}^{(\ell,m,\varepsilon,p)}\) is the additional approximation introduced by the finite adaptive HB representation.
The reconstruction tolerance \(\varepsilon\) is a numerical diagnostic and is not, by itself, an a priori bound on \(E_{\rm HB}\).

For the following comparison, we measure the errors of the same covariance restricted dynamics in either covariance geometry or
the HS norm, $g\in\{\mathrm{covar}(\sigma_0),\mathrm{HS}\}$.
Here, the label \(g\) specifies the error geometry, not a change of projection or dynamics.
For both choices, set \(P_m^{(g)}=\pi_m^{\sigma_0}\) and \(Q_m^{(g)}=1-P_m^{(g)}\), and define
\begin{align}
E_{{\rm rep},g}^{(m)}(t)
&=\normsp{Q_m^{(g)}\OH(t)}{g},
\\
E_{{\rm mem},g}^{(m)}(t)
&=\normsp{P_m^{(g)}\OH(t)-{\bf O}_m(t)}{g},
\nonumber\\
E_{{\rm HB},g}^{(\ell,m,\varepsilon,p)}(t)
&=
\normsp{
{\bf O}_m(t)-{\bf O}_{\ell,m,\varepsilon,p}(t)
}{g}.
\label{eq:si-generalized-errors}
\end{align}
For \(g=\mathrm{covar}(\sigma_0)\), these reduce to \(\Delta_{\rm rep}^{(m)}\), \(E_{\rm mem}^{(m)}\) and \(E_{\rm HB}^{(\ell,m,\varepsilon,p)}\), respectively.
A Duhamel comparison bounds the memory contribution by the representation residual of the same covariance projection.
Note that this comparison does not require the projection to be orthogonal in the norm used to measure the error.
One obtains, in a fixed norm geometry \(g\),
\begin{equation}
E_{{\rm mem},g}^{(m)}(t)
\leq L_{m,g} \int_{t_0}^{t} ds\, e^{\mu_{m,g}(t-s)} \normsp{ Q_m^{(g)}\OH(s) }{g},
\label{eq:si-memory-bound}
\end{equation}
with
\begin{equation}
L_{m,g} = \normsp{P_m^{(g)}\bm{\mathcal L}Q_m^{(g)} }{{\rm op},g}, 
\qquad
\mu_{m,g}= \normsp{P_m^{(g)}\bm{\mathcal L}P_m^{(g)}}{{\rm op},g}.
\label{eq:si-memory-constants}
\end{equation}
For full-rank reference states, the bound applies directly in covariance norm. 
For rank-deficient references, its covariance-quotient formulation additionally requires every induced map appearing in the bound to preserve covariance-null equivalence classes.
When this compatibility condition fails, as can occur in the \(m=3\) example above, the bound must instead be formulated in a positive-definite state-independent norm, or regarded as inapplicable in covariance quotient geometry.


A complementary obstruction to small representation error is provided by operator-space entanglement entropy (OSEE).
Let \(B(t)=L\cup R\) be the relevant Lieb--Robinson region and consider the operator Schmidt decomposition of \(\OH(t)\) across \(L|R\),
\begin{equation}
\OH(t)
=
\sum_i s_i\,{\bf A}_i\otimes{\bf B}_i,
\qquad
p_i
=
\frac{s_i^2}{\|\OH(t)\|_{\rm HS}^2}.
\label{eq:si-operator-schmidt}
\end{equation}
We use the R\'enyi-2 operator-space entanglement entropy
\begin{equation}
S_2(\OH;L|R) = -\log\sum_i p_i^2 ,
\label{eq:si-renyi2-osee}
\end{equation}
where, as stated above, \(\log\) denotes the natural logarithm.
Thus, $e^{S_2}=(\sum_i p_i^2)^{-1}$ which fixes the convention used in the bounds below.
If an approximation with relative squared HS error at most \(\Delta\) exists inside \({\cal A}_m\), its operator Schmidt rank must be at least the corresponding \(\Delta\)-approximate Schmidt rank.
For the \(m\)-body space this gives the necessary condition
\begin{equation}
{
(1-\Delta)^2
e^{S_2(\OH(t);L|R)}
\leq
R_m^{L|R},
}
\label{eq:si-osee-obstruction}
\end{equation}
where the cut capacity is
\begin{equation}
R_m^{L|R}
=
\min_{-1\leq k\leq m}
\left[
D_L(k)+D_R(m-k-1)
\right],
\label{eq:si-cut-capacity}
\end{equation}
with
\begin{equation}
D_X(k)
=
\sum_{a=0}^{\min(k,N_X)}
\binom{N_X}{a}3^a,
\label{eq:si-one-side-capacity}
\end{equation}
and $D_X(k)=0$ for $k<0$. 
Here \(N_L\) and \(N_R\) are the numbers of active sites on either side of the cut, such that they resolve the LR region until time $t$.
\eqref{eq:si-osee-obstruction} follows from the approximate-Schmidt-rank bound
\(
\chi_\Delta\geq(1-\Delta)^2e^{S_2}
\)
together with the maximal cut rank \(R_m^{L|R}\) available to an \(m\)-body observable.
We use \(S_2\), rather than the von Neumann OSEE, because the latter requires a dimension-dependent finite-error correction.

Defining
\begin{equation}
m_{\min,\rm rep}^{\rm HS}(t,\Delta)=
\min\left\{m:\inf_{{\bf X}\in{\cal A}_m}
\frac{\|\OH(t)-{\bf X}\|_{\rm HS}^2
}{\|\OH(t)\|_{\rm HS}^2}
\leq\Delta
\right\},
\label{eq:si-minimal-hs-order}
\end{equation}
the smallest \(m\) satisfying~\eqref{eq:si-osee-obstruction},
denoted \(M_{\rm nec}^{\rm HS}(t,\Delta)\), obeys
\begin{equation}
{
M_{\rm nec}^{\rm HS}(t,\Delta)
\leq
m_{\min,\rm rep}^{\rm HS}(t,\Delta).
}
\label{eq:si-osee-lower-bound}
\end{equation}
OSEE, therefore, provides a necessary representational obstruction, not a sufficient accuracy criterion and not a direct estimator of the body order required for state-adapted expectation values.
In particular,~\eqref{eq:si-osee-lower-bound} is a state-independent HS statement, whereas the covariance projection may regard large HS components as statistically weak for the chosen \(\sigma_0\).
The full derivation of the OSEE-obstruction is reserved for the accompanying work \cite{Perez_long_paper_2026}.

A sufficient bound requires additional information about the distribution of operator weight over body order.
For the sufficient-bound derivation below, we take \(g=\mathrm{covar}(\sigma_0)\) and retain the same projection used in the dynamics, \(\mathsf P_r^{(g)}=\pi_r^{\sigma_0}\).
We define the body-sector projections by
\begin{equation}
\Pi_0^{(g)}=\mathsf P_0^{(g)}, \quad
\Pi_r^{(g)} =\mathsf P_r^{(g)}-\mathsf P_{r-1}^{(g)}, \,\,\, r\geq1,
\end{equation}
and their weights by
\begin{equation}
\Theta_r^{(g)}(t) = \normsp{\Pi_r^{(g)}\OH(t)}{g}^2.
\label{eq:si-body-sector-weight}
\end{equation}
Covariance orthogonality of the centered body sectors gives the exact identity
\begin{equation}
\bigl[E_{{\rm rep},g}^{(m)}(t)\bigr]^2 =
\normsp{(1-\mathsf P_m^{(g)})\OH(t)}{g}^2 =
\sum_{r>m}\Theta_r^{(g)}(t).
\label{eq:si-body-tail-identity}
\end{equation}
This identity holds as a seminorm identity for rank-deficient product references as well.
Importantly, it does not require covariance orthogonality between individual strings within a given support or body-order sector.
Nevertheless, its use in the dynamical memory bound remains subject to the null-space compatibility conditions stated above.

For a normalized Heisenberg-evolved Pauli observable, $\|\OH(t)\|_{\rm HS}=1$ with
$$
\OH(t)=\sum_P c_P(t)P
$$,
define
\begin{equation}
{\cal M}_1(t) = \log\!\left(
\sum_P |c_P(t)| \right).
\label{eq:si-pauli-l1}
\end{equation}
This logarithmic Pauli-\(\ell_1\) quantity is proportional to the \(\alpha=\frac12\) operator stabilizer R\'enyi entropy of \cite{dowling2025_magic}.
That reference employs base-\(2\) logarithms, whereas we use natural logarithms throughout this section; consequently,
\begin{equation}
{\cal M}_1 = \frac{\ln 2}{2}\,
{\cal M}^{(1/2)}
\label{eq:si-magic-convention}
\end{equation}
in the corresponding normalized-Pauli convention.
Its global value alone does not determine the body-order distribution.

We therefore impose the explicit concentration hypothesis
\begin{equation}
\Theta_r^{(g)}(s)
\leq
C_g
e^{2{\cal M}_1(s)}
e^{-2\kappa_g r},
\qquad
0\leq s\leq t,
\label{eq:si-tail-assumption}
\end{equation}
with \(C_g,\kappa_g>0\).
The exponential factor in \(r\) is an independent body-order concentration assumption and is not implied by \({\cal M}_1\).

Summing~\eqref{eq:si-body-tail-identity} under~\eqref{eq:si-tail-assumption} gives
\begin{equation}
E_{{\rm rep},g}^{(m)}(t) \leq A_{\kappa,g} e^{{\cal M}_1(t)} e^{-\kappa_g(m+1)},
\label{eq:si-representation-upper}
\end{equation}
where $A_{\kappa,g} = \sqrt{
\frac{C_g}{1-e^{-2\kappa_g}}
}$. 
Hence, a sufficient condition for $E_{{\rm rep},g}^{(m)}(t)\leq\eta_{\rm rep}$ is
\begin{equation}
{
m \geq \frac{1}{ \kappa_g} \bigg[{ {\cal M}_1(t) + \log[A_{\kappa,g}/\eta_{\rm rep}] } \bigg] -1 .
}
\label{eq:si-representation-threshold}
\end{equation}
The smallest integer satisfying~\eqref{eq:si-representation-threshold} is, therefore, a conditionally certified feasible body order and an upper bound on the corresponding minimal representation order.

The same concentration hypothesis also bounds the memory contribution.
With
\begin{equation}
{\cal M}_1^\star(t) = \sup_{t_0\leq s\leq t} {\cal M}_1(s),
\label{eq:si-Mstar}
\end{equation}
\eqref{eq:si-memory-bound} and~\eqref{eq:si-representation-upper} imply
\begin{multline}
E_{\rm mem, g}^{(m)}(t)
\leq
A_{\kappa,g}L_{m, g}(t-t_0) \\ 
\times \exp\bigg({{\cal M}_1^\star(t)
+\mu_{m,g}(t-t_0)-\kappa_g(m+1)\bigg)}.
\label{eq:si-memory-upper}
\end{multline}
Consequently,
\begin{equation}
{
m \geq \frac{1}{\kappa_g}\bigg[ {\cal M}_1^\star(t)
+\mu_{m,g} \tau + \log[
A_{\kappa,g}L_{m,g} \tau/\eta_{\rm mem}]
\bigg] -1
}
\label{eq:si-memory-threshold}
\end{equation}
is sufficient for $E_{{\rm mem},g}^{(m)}(t)\leq\eta_{\rm mem}$ with $\tau=t-t_0$.
Because \(L_{m,g}\) and \(\mu_{m,g}\) themselves depend on \(m\), this condition defines an implicit certified set of body orders rather than an asymptotic scaling law. 
Finally, assigning an error budget
\begin{equation}
\eta_{\rm rep} +
\eta_{\rm mem} +
\eta_{\rm HB} \leq
\delta,
\label{eq:si-error-budget}
\end{equation}
~\eqref{eq:si-representation-threshold} and~\eqref{eq:si-memory-threshold}, together with an independently established $E_{\rm HB, g}^{(\ell,m,\varepsilon,p)} \leq\eta_{\rm HB}$ provide a sufficient condition for the prescribed total accuracy.
The resulting upper bound and the OSEE obstruction bracket the same minimal body order only when both are formulated in the same HS geometry.
When the tail hypothesis~\eqref{eq:si-tail-assumption} is imposed in covariance geometry, it instead gives a state-adapted sufficient upper bound, while~\eqref{eq:si-osee-lower-bound} remains a distinct state-independent HS obstruction.

Thus, OSEE and nonstabilizerness play complementary but fundamentally different roles with the OSEE constraining the minimum Schmidt-rank capacity that any low-body HS-representation must possess, whereas \({\cal M}_1\) enters a constructive sufficient bound only together with an independently controlled body-order tail.
The sufficient bounds conditionally control errors, in the covariance norm, of the restricted dynamics, whereas \eqref{eq:si-osee-lower-bound} is a state-independent obstruction
to low-body representation in HS norm.
Neither quantity alone determines \(m(t,\delta)\) nor, as explained previously, yields tight bounds on the analogous state-adapted covariance seminorms, which may very well be much lower and tighter than any HS-bound.
Further derivations and questions of sharpness are discussed in the companion work \cite{Perez_long_paper_2026}.

\paragraph{Implementation details}

All numerical benchmarks were performed on the same workstation equipped with an AMD Ryzen 9 5900X CPU, comprising 12 physical cores and 24 hardware threads, and \(64\,\mathrm{GB}\) of RAM.
Unless stated otherwise, reported wall times correspond to single-threaded CPU execution without GPU acceleration.
The PP implementation is separated into a structural precomputation stage, in which the admissible at-most-\(m\)-body Pauli sector and the corresponding projected operator algebra are constructed, and an online stage in which the adaptive HB dynamics of~\eqref{eq:hb_ansatz}--\eqref{eq:hb_galerkin} are propagated for the chosen \(\h\) and \(\sigma_0\), following the workflow of Refs.~\cite{Vetter_2026,Perez_long_paper_2026}.
At fixed \(m\), the covariance and HS calculations therefore share a substantial part of the structural workload, while their online costs differ through the Pauli content generated and retained by the respective projection rules.
Unless stated otherwise, end-to-end wall times explicitly include both precomputation and online evolution and are cold started end-to-end measurements, with no overlap between different simulations.
Each timing corresponds to a single measurement. 

Pauli coefficients with magnitude below \(\varepsilon_{\mathrm{count}}=\epsilon_{\rm nnz}=10^{-12}\) are treated as zero for support-size and sparsity diagnostics, while expectation-value contractions use \(\varepsilon_{\mathrm{expect}}=10^{-15}\).
Independently, HB contributions whose coordinate magnitude is below \(10^{-12}\) are omitted during operator assembly as a floating-point cleanup threshold.

The formal reduced dynamics is written using the Moore--Penrose prescription in~\eqref{eq:hb_coeff_eom}.
For a compatible rank-deficient chart, this prescription selects the minimum-norm resolved coefficient evolution.
The production implementation in \textsc{Phoenix}, however, does not explicitly form the Moore--Penrose pseudoinverse.
Instead, once per HB chart, the covariance Gram matrix is symmetrized according to
\begin{equation}
{\cal G} \leftarrow \frac{1}{2}
\left({\cal G}+{\cal G}^{T}\right).
\end{equation}
A weak ridge regularization is then introduced as
\begin{equation}
r=\varepsilon_{\mathrm{G}} \max\!\left(
\|{\cal G}\|_{2}, 1\right), \end{equation}
where \(\varepsilon_{\mathrm{G}}\) is the Gram-regularization parameter.
For the production calculations, we use \(\varepsilon_{\mathrm G}=10^{-12}\).
The resulting ridge \(r\) varies from chart to chart through the scale \(\max(\|{\cal G}\|_2,1)\).
The regularized matrix
\begin{equation}
{\cal G}_{r} = {\cal G}+rI
\end{equation}
is factorized using LU decomposition, and the resulting factors are cached and reused for all Gram solves within that HB chart.
The numerical coefficient-space generator is therefore
\begin{equation}
{\cal A}_{r} =\left(
{\cal G}+rI \right)^{-1}
{\cal H}.
\end{equation}
This ridge regularization acts only on the finite-dimensional coordinate solve and does not alter the state-adapted projection \(\pi_m^{\sigma_0}\), the retained body order \(m\), or the projected operator algebra.
At finite \(r\), it should nevertheless be distinguished from the formal Moore--Penrose solution, since it regularizes the coefficient-space generator itself.

The TN calculations were performed independently with \textsc{TeNPy}~\cite{tenpy} using Krylov-enriched two-site TDVP--MPO evolution.
Accordingly, the reported wall times are end-to-end measurements of the present implementations and should not be interpreted as hardware-independent complexity bounds or intrinsic speed ratios between PP and TN methods.
The magnetization \(Z_0(t)\) is evaluated on the uniform grid \(t_kJ=0.005k\) for \(k=0,\ldots,243\).
We use maximum bond dimensions \(\chi_{\max}=8,16,32,64,\) and \(128\), with a common singular-value threshold \(\varepsilon_{\mathrm{SVD}}=10^{-9}\).
The reported \(\chi_{\max}\) is an upper bound, because bonds are not artificially enlarged when their numerical rank is smaller.
The initial operator \(Z_0\) has bond dimension one, but its second derivative \(\mathsf{K}|Z_0\rangle\rangle\) already contains interaction-generated components that can cross distant cuts of the ``snake'' representation  to one-dimensional TN ansatz and, as an undesirable consequence, local two-site updates need not expose all of these directions to the retained virtual bases sufficiently early.
Increasing \(\chi_{\max}\) allows more singular directions to survive once generated, but does not by itself determine which directions are initially proposed to the variational subspace.
In unenriched calculations, this produces an incorrect short-time dip and an apparent convergence within an inadequately populated tangent-space representation.
This failure to converge towards the exact solution in the \(N=18\) example disappears when utilizing \(d_{\rm Krylov}=1\).
For more details, refer to \cite{Perez_long_paper_2026}.

As an early-time diagnostic, we also tested \(\varepsilon_{\mathrm{SVD}}=10^{-14}\) for \(tJ\leq0.3\) before returning to \(10^{-9}\).
Its agreement with the enriched calculation supports the interpretation in terms of early rank starvation, although the two remedies are not entirely equivalent.
The production trajectories use the uniform threshold \(10^{-9}\) together with one-vector enrichment.

\paragraph{Complementary analysis of the \(N=18\) benchmark}

Previously, in the main results section, we have characterized the \(3\times3\times2\), \(N=18\) TFIM benchmark through the instantaneous error and the corresponding accuracy--wall-time tradeoff in Fig.~\ref{fig:covariance-advantage-single}.
Here, we complement that discussion by displaying the dynamics of $Z_0(t)_{\rm exact}$ itself, the different PP and TN approximations, and by comparing the maximum error with the realized sparse size of the Pauli-propagation representation.

\begin{figure}[t]
\centering
\includegraphics[width=1\linewidth]{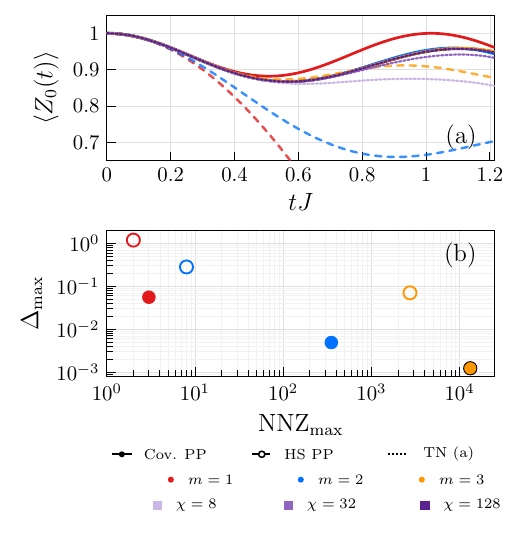}\caption{
\textbf{Observable dynamics and sparse-representation size for the \(3\times3\times2\) TFIM benchmark.}
(a) Corner polarization \(\langle Z_0(t)\rangle\) for the \(N=18\) system of Fig.~\ref{fig:covariance-advantage-single}
, with \(h/J=0.75\), \(J=1\), and the fully \(z\)-polarized initial state.
The exact result is shown in black, solid and dashed curves denote covariance- and HS-restricted adaptive Pauli propagation at \(m=1,2,3\), respectively, and dotted curves show representative Krylov-enriched TDVP--MPO calculations at \(\chi=8,32,128\).
(b) Maximum instantaneous error \(\Delta_{\max}\) versus the maximum active-basis sparse size \(\mathrm{NNZ}_{\max}\) reached by each Pauli-propagation calculation.
Filled and open circles denote covariance and HS propagation, respectively.
Tensor-network data are not shown in panel (b), since a Pauli-string NNZ is not defined for the MPO representation.
}
\label{fig:si-tfim-dynamics-nnz}
\end{figure}

Figure~\ref{fig:si-tfim-dynamics-nnz}(a) shows directly, on the scale of the physical observable, the behavior summarized by the error curves of Fig.~\ref{fig:covariance-advantage-single}(a).
The HS hierarchy improves systematically as \(m\) is increased, but substantial deviations remain at fixed body order and breakdowns of validity are clear and apparent. 
At \(m=1\), the HS trajectory departs rapidly from the exact evolution and does not reproduce the subsequent revival, while covariance \(m=1\) retains the qualitative structure of the dynamics throughout the displayed interval.
At \(m=2\), the covariance trajectory closely follows both the first dip in polarization and the subsequent revival, whereas the HS trajectory develops a sizeable systematic deviation at intermediate times.
At this retained body order $m=2$, covariance dynamics is indistinguishable from the exact dynamics in panel (a).
Furthermore, at \(m=3\), the covariance result is also indistinguishable from the exact curve on the scale of panel (a) and from its $m=2$ covariance predecessor, consistent with the \(10^{-3}\)-scale maximum error reported in the main text.
The independent TDVP--MPO hierarchy approaches the same exact trajectory as \(\chi\) is increased, providing a separate convergence sequence based on a fundamentally different representation and propagation algorithm. 

To quantify how much sparse Pauli structure is required to achieve this accuracy, we use
\begin{equation}
\Delta_{\max}
=
\max_{0\leq tJ\leq1.215}
\left|
\langle Z_0(t)\rangle_{\rm approx}
-
\langle Z_0(t)\rangle_{\rm exact}
\right|,
\label{eq:si-max-error-nnz}
\end{equation}
together with the maximum number of nonzero Pauli coefficients appearing in any active HB basis operator,
\begin{equation}
\mathrm{NNZ}_{\max}
=
\max_{n,\alpha}
\operatorname{nnz}\!\left(
{\bf b}^{(n)}_\alpha
\right),
\label{eq:si-hb-instantaneous-nnz}
\end{equation}
with $n$ associated the  reconstruction times $T_n$.
As in the main text, coefficients with magnitude below \(\epsilon_{\rm nnz}=10^{-12}\) are treated as zero so as to avoid a deceptive nnz counting a large number of near-zero observable components present in the ansatz ${\bf O}_{\ell, m,\varepsilon,p}(t)$.
Unlike the formal cutoff \(m\), which specifies the largest retained body order, \(\mathrm{NNZ}_{\max}\) measures how much of that admissible sparse sector is actually populated by the adaptive calculation.

Panel~\ref{fig:si-tfim-dynamics-nnz}(b) shows that the covariance advantage cannot be reduced to simply retaining more Pauli strings.
At \(m=1\), both geometries remain extremely sparse, yet the covariance geometry lowers the maximum error from order unity to the \(10^{-2}\) scale, fundamentally indicating that, at this body order $m=1$, the HS- and covariance adaptive ans\"atze are obtained from completely different criteria.
The improvement, therefore, genuinely arises primarily from the information encoded in the compressed low-body representative rather than from a substantial increase in representation size.

At \(m=2\), the covariance calculation reaches a maximum error of a few \(10^{-3}\), while HS remains at the \(10^{-1}\) scale.
The covariance representative is also appreciably denser, showing that beyond the lowest order the state-adapted projection both changes the retained coefficients and populates a larger fraction of the admissible \({\cal A}_m\) sector.
The additional sparse content accompanies an improvement in accuracy approaching two orders of magnitude.

The same tendency becomes more pronounced at \(m=3\) with the covariance calculation reaching the \(10^{-3}\) error scale with an active representation containing approximately an order of magnitude more Pauli coefficients than its HS counterpart, while the HS error remains at the \(10^{-2}\) scale.
Note that a less sparse, denser, ansatz is correlated with a higher end-to-end wall time, as reported in Fig.~\ref{fig:covariance-advantage-single}.
This behavior is consistent with the action of \(\pi_m^{\sigma_0}\), components generated outside of ${\cal A}_m$ can contribute nontrivial state-relevant combinations inside this space, whereas HS weight truncation removes those high-body components without such compression.

The larger \(\mathrm{NNZ}_{\max}\) at higher covariance orders should therefore not be interpreted as the sole reason for a covariance advantage over HS dynamics.
Rather, Fig.~\ref{fig:si-tfim-dynamics-nnz}(b) separates two effects of the state-adapted construction.
At low body order, a substantial accuracy improvement is obtained with almost no increase in explicit sparse size simply by redefining what the ansatz is going to be according to a state-relevant criterion, whereas at larger \(m\) further accuracy is accompanied by denser representatives within the same formal at-most-\(m\) sector.
Finally, \(\mathrm{NNZ}_{\max}\) is specific to the sparse Pauli representation and is not used as a direct representation-size comparison with the TN MPO calculations.

\paragraph{Complementary cost analysis of the \(N=90\) benchmark}

The \(6\times5\times3\), \(N=90\) benchmark of Fig.~\ref{fig:scaling-nnz-inset} compares the observable dynamics with the instantaneous sparse content generated at the HB reconstruction points.

Here, we complement the analysis of the linear fits of ${\rm NNZ}(T_n; \epsilon_{\rm nnz})$ vs. $(T_nJ)^\alpha$ with a more complete analysis of the stability of said fits. 
The main text characterizes the finite-time growth of the instantaneous NNZ through~\eqref{eq:hb-instantaneous-nnz}.
Table~\ref{tab:scaling-fits-si} gives the complete fit diagnostics underlying the effective exponents quoted there, including the fitted time window, number of reconstruction points, standard errors, confidence intervals, and sensitivity to removing the earliest or latest fitted point.

\begin{table*}[t]
\caption{
Finite-time fits of the last-HB-vector sparsity across different HBs, $\mathrm{NNZ}(b_{\rm last}^{(n)})\propto(tJ)^\alpha$, for the $d=3$, $6\times5\times3$ TFIM benchmark.
$N_{\rm fit}$ is the number of positive-time HB reconstruction points entering the log--log regression. 
Uncertainties are standard errors and brackets give 95\% confidence intervals. 
The endpoint-sensitivity exponents $\alpha_{-\rm early}$ and $\alpha_{-\rm late}$ are obtained after omitting the earliest or latest fitted point, respectively and quantify sensitivity of the finite-time fit and should not be interpreted as separate early- or late-time scaling exponents. 
Constant entries exhibit no resolved NNZ growth over the sampled interval.
The instantaneous worst-case exponent is $\alpha_{\rm wc}=dm$ as reported in \eqref{eq:complexity}.
}
\label{tab:scaling-fits-si}
\centering
\scriptsize
\setlength{\tabcolsep}{4.5pt}
\begin{ruledtabular}
\begin{tabular}{ccccccccc}
geometry
&
$m$
&
$tJ$ range
&
$N_{\rm fit}$
&
$\alpha\pm{\rm SE}$
&
95\% CI
&
$\alpha_{-\rm early}$
&
$\alpha_{-\rm late}$
&
$\alpha_{\rm wc}$
\\
\colrule
cov.
&
$1$
&
$0.015$--$1.215$
&
$81$
&
$0$
&
--
&
$0$
&
$0$
&
$3$
\\
cov.
&
$2$
&
$0.135$--$1.215$
&
$9$
&
$1.105\pm0.063$
&
$[0.957,1.253]$
&
$0.954$
&
$1.142$
&
$6$
\\
cov.
&
$3$
&
$0.135$--$1.080$
&
$8$
&
$2.319\pm0.115$
&
$[2.037,2.602]$
&
$2.082$
&
$2.401$
&
$9$
\\
HS
&
$1$
&
$0.015$--$1.215$
&
$81$
&
$0$
&
--
&
$0$
&
$0$
&
$3$
\\
HS
&
$2$
&
$0.015$--$1.215$
&
$81$
&
$0$
&
--
&
$0$
&
$0$
&
$6$
\\
HS
&
$3$
&
$0.135$--$1.215$
&
$9$
&
$1.603\pm0.103$
&
$[1.359,1.847]$
&
$1.304$
&
$1.624$
&
$9$
\\
\end{tabular}
\end{ruledtabular}
\end{table*}

The fits in Table~\ref{tab:scaling-fits-si} remain well below the corresponding instantaneous LR retained-space estimates of \eqref{eq:complexity}.
For covariance dynamics, the fitted exponents are \(\alpha=1.105\pm0.063\) at \(m=2\) and \(\alpha=2.319\pm0.115\) at \(m=3\), compared with \(\alpha_{\rm wc}=6\) and \(9\), respectively.
For HS PP, no NNZ growth is resolved at \(m=2\), while \(m=3\) gives \(\alpha=1.603\pm0.103\).

The endpoint tests also quantify the finite-window character of these exponents.
For covariance \(m=2\), removing the earliest or latest fitted point gives \(\alpha=0.954\) and \(1.142\), respectively, compared with \(\alpha=1.105\) for the complete fit.
For covariance \(m=3\), the corresponding values are \(2.082\) and \(2.401\), compared with \(2.319\), while HS \(m=3\) gives \(1.304\) and \(1.624\), compared with \(1.603\).
The observed variation reinforces that these quantities should be regarded as finite-time effective growth exponents rather than asymptotic power laws.
Note that for covariance $m=3$, the fitted window ends at $T_n J = 1.080$ since that is the last reconstruction time. 

\begin{figure}[t]
\centering
\includegraphics[width=1\linewidth]{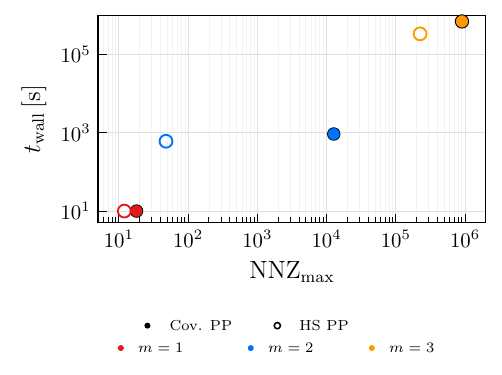}\caption{
\textbf{Computational cost and sparse-representation size for the \(N=90\) TFIM benchmark.}
Measured end-to-end wall time \(t_{\rm wall}\) versus the maximum active-basis sparse size \(\mathrm{NNZ}_{\max}\) reached by covariance and HS adaptive PP at \(m=1,2,3\).
Filled and open circles denote covariance and HS PP, respectively.
Data correspond to the \(6\times5\times3\) benchmark of Fig.~\ref{fig:scaling-nnz-inset}, with \(\ell=5\), \(p=2\), and \(\varepsilon=10^{-10}\).
Tensor-network data are not included because a Pauli-basis NNZ is not defined for the MPO representation.
Both axes are logarithmic.
}
\label{fig:si-n90-walltime-nnz}
\end{figure}

Taken together, Fig.~\ref{fig:si-n90-walltime-nnz} and Table~\ref{tab:scaling-fits-si} separate two complementary notions of numerical complexity.
The former relates the largest realized sparse representation to the measured end-to-end computational cost, while the latter characterizes how the instantaneous sparse content develops over the simulated time window.
Neither quantity alone determines the quality of the approximation, which is instead assessed through the finite-\(m\) dynamical hierarchy in the main discussion.

We complement this analysis by displaying a complementary discussion relating the maximum active-basis sparse size, $\rm NNZ_{\rm max}$, reached by each PP calculation to its measured end-to-end wall time.
This comparison separates the representation-level sparsity characterized in the main text from the implementation-dependent computational effort required to propagate it.

Figure~\ref{fig:si-n90-walltime-nnz} shows that the realized sparse size and the measured computational cost are related but are not interchangeable quantities.
For \(m=1\), both geometries remain extremely sparse, with \(\mathrm{NNZ}_{\max}=18\) for covariance PP and \(12\) for HS PP, while both calculations require approximately \(10\,{\rm s}\).

The distinction is considerably stronger at \(m=2\).
Covariance PP reaches \(\mathrm{NNZ}_{\max}=12714\), compared with only \(48\) for HS PP, while HS dynamics require \(604\,{\rm s}\) and covariance dynamics \(925\,{\rm s}\), approximately \(53\%\) longer.
Thus, a difference of more than two orders of magnitude in the maximum sparse size does not produce a comparable difference in end-to-end wall time.
The latter also contains the structural construction, projected commutators, HB reconstructions, Gram and coordinate-space operations, and the adaptive evolution itself, and therefore cannot be inferred from the maximum active-basis NNZ alone.

At \(m=3\), the larger representation is accompanied by a corresponding rather large increase in computational effort.
Covariance PP reaches \(\mathrm{NNZ}_{\max}=902093\) and \(t_{\rm wall}=688366\,{\rm s}\), around 200 hours, compared with \(\mathrm{NNZ}_{\max}=224160\) and \(t_{\rm wall}=332317\,{\rm s}\) for HS PP, which is approximately 100 hours.
The covariance representation is therefore approximately four times denser at its largest active HB operator and approximately twice as expensive in end-to-end wall time in the present implementation.
As established from the dynamics in Fig.~\ref{fig:scaling-nnz-inset}, this additional representation and computational cost accompanies the substantially stronger agreement between successive covariance body orders.

Figure~\ref{fig:si-n90-walltime-nnz} should therefore not be interpreted as defining NNZ as a computational-cost measure.
Rather, it shows explicitly how the sparse representation realized by the two projection geometries relates, nontrivially, to the measured cost of the corresponding simulations.
The wall times remain implementation- and hardware-dependent, whereas \(\mathrm{NNZ}_{\max}\) characterizes the largest realized sparse PP representation independently of those details.
\end{document}